\documentclass[onecolumn,prl, preprint, superscriptaddress]{revtex4-2}
\usepackage{bm}
\usepackage[colorlinks=true,linkcolor=blue,citecolor=blue]{hyperref}
\usepackage{times}
\usepackage{amsmath}
\usepackage{amssymb}
\usepackage{amsthm}
\usepackage{amsfonts}
\usepackage{enumerate}
\usepackage{latexsym}
\usepackage{ifpdf}
\usepackage{natbib}
\usepackage{psfrag}
\newcommand{\beq}{\begin{equation}}
\newcommand{\eeq}{\end{equation}}
\usepackage{graphicx}
\usepackage{makeidx}
\usepackage{color}
\usepackage{array}
\usepackage{multirow}
\usepackage{siunitx}
\begin{document}
	
	\title{Orthorhombic distortion drives orbital ordering in an antiferromagnetic 3$d^1$ Mott insulator}
	\author {Prithwijit Mandal}
	\email{prithwijitm@iisc.ac.in}
	\affiliation  {Department of Physics, Indian Institute of Science, Bengaluru  560012, India}
	\author {Shashank Kumar Ojha}
	\affiliation  {Department of Physics, Indian Institute of Science, Bengaluru 560012, India}
	\author{Duo Wang}
	\affiliation{Department of Physics and Astronomy, Uppsala University,  Uppsala 75120, Sweden}
	\affiliation{Faculty of Applied Sciences, Macao Polytechnic University, Macao, China}
	\author {Ranjan Kumar Patel}
	\affiliation  {Department of Physics, Indian Institute of Science, Bengaluru 560012, India}
	\author {Siddharth Kumar}
	\affiliation  {Department of Physics, Indian Institute of Science, Bengaluru  560012, India}
	\author {Jyotirmay Maity}
	\affiliation  {Department of Physics, Indian Institute of Science, Bengaluru  560012, India}
	\author{Zhan Zhang}
	\affiliation {Advanced Photon Source, Argonne National Laboratory, Lemont, Illinois 60439, USA}
	\author{Hua Zhou}
	\affiliation {Advanced Photon Source, Argonne National Laboratory, Lemont, Illinois 60439, USA}
	\author{Christoph Klewe}
	\affiliation  {Advanced Light Source, Lawrence Berkeley National Laboratory, Berkeley, California 94720, USA}
	\author {Padraic Shafer}
	\affiliation  {Advanced Light Source, Lawrence Berkeley National Laboratory, Berkeley, California 94720, USA}
	\author{Biplab Sanyal}
	\email{biplab.sanyal@physics.uu.se}
	\affiliation{Department of Physics and Astronomy, Uppsala University,  Uppsala 75120, Sweden}
	\author {Srimanta Middey}
	\email{smiddey@iisc.ac.in}
	\affiliation  {Department of Physics, Indian Institute of Science, Bengaluru 560012, India}
	
	\begin{abstract}
		{ \textbf{The orbital, which represents the shape of the electron cloud, very often strongly influences the manifestation of various exotic phenomena, e.g., magnetism, metal-insulator transition, colossal magnetoresistance, unconventional superconductivity etc. in solid-state systems.  The observation of the antiferromagnetism in $\textbf{\textit{RE}}$TiO$_3$ ($\textbf{\textit{RE}}$=rare earth) series has been puzzling since the celebrated Kugel-Khomskii model of spin-orbital super exchange predicts ferromagnetism in an orbitally degenerate $\textbf{\textit{d}}^\textbf{1}$ systems. Further, the existence of the orbitally ordered vs. orbital liquid phase in both antiferromagnetic and paramagnetic phase have been unsettled issues thus far.  To address these long-standing questions, we investigate single crystalline film of PrTiO$_3$. Our synchrotron X-ray diffraction measurements confirm the retention of bulk-like orthorhombic ($\textbf{\textit{D}}_\textbf{2h}$) symmetry in the thin film geometry. We observe similar X-ray linear dichroism signal in both paramagnetic and antiferromagnetic phase, which can be accounted by ferro orbital ordering (FOO). While the presence of $\textbf{\textit{D}}_\textbf{2h}$ crystal field does not guarantee lifting of orbital degeneracy always, we find it to be strong enough in these rare-earth titanates, leading to the FOO state. Thus, our work demonstrates the orthorhombic distortion is the driving force for the orbital ordering of antiferromagnetic $\textbf{\textit{RE}}$TiO$_3$.}}
		\end{abstract}
	
	\maketitle
	
	\section*{Introduction}	
	
	The shape of the quantum mechanical wave function for an electron bound to an atomic nucleus by Coulomb force is determined by the orbital ($l$) and magnetic quantum ($m_l$) numbers. When different atoms/ions are combined to form a solid, the azimuthal degeneracy is lifted due to the presence of Coulomb interaction with the neighboring atoms/ions. For example, in transition metal oxides (TMOs) with octahedral crystal field (CF), the original five-fold degenerate $d$ orbitals [$l$=2; $m_l$=0, $\pm$1, $\pm$2] of the TM site are split into three fold degenerate $t_{2g}$ orbitals with lower energy and doubly degenerate $e_{g}$ orbitals with higher energy~\cite{Griffith:1964book}. Very often, such degeneracy is further lifted by structural distortion, leading to an orbital ordering (OO)~\cite{Khomskii:2022p054004}.  Interestingly, OO can also happen in absence of any CF splitting   due to electronic superexchange, popularly known as   Kugel-Khomskii mechanism~\cite{Kugel:1973p1429}. In reality, both of these mechanisms of OO can coexist and separating the role of electron-lattice vs. spin-orbital coupling is nontrivial due to their comparable energy scale~\cite{Khomskii:2014book}. The presence of OO and its role in magnetism, metal-insulator transition, superconductivity, charge density wave phase, structural symmetry change etc. has been demonstrated in various complex oxides (see Ref.~\cite{Tokura:2000p462,Streltsov:2017p1121,Khomskii:2022p054004} for review).

Mott insulating rare-earth titanates ($RE$TiO$_3$) exhibit an interesting phase diagram with a crossover from antiferromagnetic to ferromagnetic phase as a function of Ti-O-Ti bond angle~\cite{Zhou:2005p7395}.
 The origin of  ferromagnetic ordering in  highly distorted members like YTiO$_3$ is well understood considering  Kugel-Khomskii coupling between  spin and orbital degrees of freedom with two-fold degenerate lower $t_{2g}$ levels~\cite{Nakao:2002p184419,Kugel:1973p1429,Mochizuki:2004p154}.
	On the contrary, the appearance of $G$-type antiferromagnetic phase for the less distorted members (bulk) e.g. LaTiO$_3$ (LTO)  remains a subject of intense scrutiny without reaching any consensus~\cite{Mizokawa:1996p5368,Khaliullin:2000p3950,Mochizuki:2001p2872,Mochizuki:2003p167203,Mochizuki:2003p167203,Mochizuki:2004p154,Pavarini:2004p176403,Pavarini:2005p188,Varignon:2017p235106}.
	Antiferromagnetic LaTiO$_3$ (LTO) has only a small distortion of its TiO$_6$ octahedra ~\cite{Maclean:1979p35} and  the $t_{2g}$ levels were initially considered to be degenerate. This  would result in  ferromagnetism in $d^1$ system according to the  Kugel-Khomskii model~\cite{Kugel:1973p1429}.
	
	The observation of a small moment  on Ti for LTO  was   thought  to be an outcome of sizable spin orbit coupling~\cite{Meijer:1999p11832}. However, such scenario was discarded by the observation of an isotropic spin wave spectra~\cite{Keimer:2000p3946}. Also,  resonant X-ray scattering study could not reveal any sizable order parameter for orbital ordering~\cite{Keimer:2000p3946}. These results were explained by a quantum fluctuation-driven orbital liquid (OL) model~\cite{Khaliullin:2000p3950}. Contrary to this, the presence of rigid ferro orbital ordering (FOO) was proposed considering the degeneracy lifting of $t_{2g}$ states due to the antiparallel displacement of $RE$ ions~\cite{Mochizuki:2003p167203}. Such a scenario   was further promoted by several theoretical and experimental studies~\cite{Pavarini:2004p176403,Cwik:2003p060401,Kiyama:2003p167202,Haverkort:2005p056401}. Overall, the puzzle about the actual orbital scenario of antiferromagnetic $RE$TiO$_3$ is yet to be solved.

	\begin{figure}
		\vspace{-0pt}
		\includegraphics[width=0.80\textwidth] {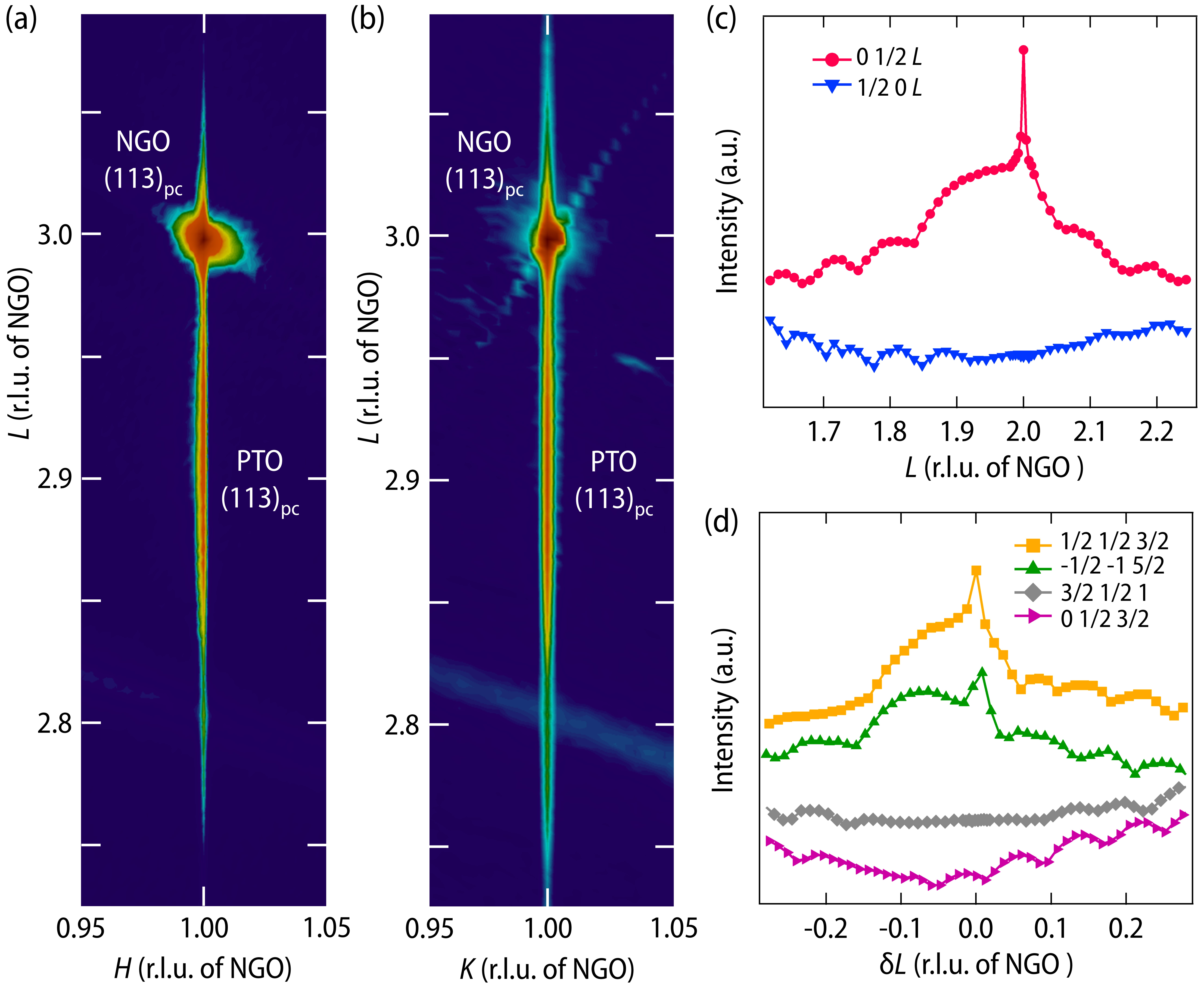}
		\caption{\label{Fig1} {(a) and (b) shows the reciprocal space mappings onto the $H$-$L$ and $K$-$L$ plane respectively. The intensity of the (1 1 3)$_\mathrm{pc}$ diffraction spots, from both the substrate and the film, peak at same $H$ [in (a)] and $K$ [in (b)] values, signifying the coherently strained nature of the film. Thus,  in-plane lattice constants of the film are $a_\mathrm{IP}$ = 3.863 $\mathrm{\AA}$, $b_\mathrm{IP}$ = 3.854 $\mathrm{\AA}$ [see Supplemental Material (SM)~\cite{SM}]. Elongation of the out of plane lattice constant $c_\mathrm{OP}$ $\sim$ 4.011 \AA\ compared to the bulk PTO  arises due to the compressive strain from the substrate. (c) $L$-scan around the (0 1/2 2)$_\mathrm{pc}$ and (1/2 0 2)$_\mathrm{pc}$ crystal truncation rods. (d) Octahedral rotational pattern has been determined by  $L$-scan centered around different half-order diffraction ($H' K' L'$) spots ($\delta L$ = $L - L'$).}}

	\end{figure}

	\section*{Results}	
	
	Here, we have explored the multiorbital physics of antiferromagnetic $RE$TiO$_3$ by  X-ray linear dichroism spectroscopy (XLD)~\cite{Chakhalian:2007p1114}. We find that the isotropic X-ray absorption spectroscopy fails to differentiate among different symmetries of the crystal field for $RE$TiO$_3$, and XLD experiments play a pivotal role to settle  the debate between FOO and OL state. Our results clarify that the level splitting due to $D_{2h}$ symmetry is strong enough to favor OO, which is further corroborated by density functional theory. We also find the OO pattern is independent of the magnetic transition, resolving another unsettled issue about the role of antiferromagnetism in setting OO in $RE$TiO$_3$ ~\cite{Haverkort:2005p056401,Komarek:2007p224402,Zhang:2020p035113}.

	To examine orbital physics, we have investigated thin epitaxial PrTiO$_3$ (PTO) films [thickness: 10-20 uc] grown on   NdGaO$_3$ (NGO) (1 1 0)$_\mathrm{or}\equiv$(0 0 1)$_\mathrm{pc}$ substrates by pulsed laser deposition [$\mathrm{or}$, $\mathrm{pc}$ denotes  orthorhombic, pseudocubic setting, also see SM~\cite{SM}]. Details of sample characterizations have been shown in SM~\cite{SM}. We have carefully chosen this compound  as the bulk PTO has a lower Ti-O-Ti bond angle compared to LTO, which would enhance the GdFeO$_3$ distortion. Also, the higher Mott gap makes PTO a robust insulator - unlike LTO, which easily becomes metallic by excess oxygen doping ~\cite{Scheiderer:2018p1706708}.
	The  resistivity ($\rho$)  [see SM ~\cite{SM}] exhibits activated behavior ($\rho$ =$\rho_0$ exp($E_\mathrm{a}$/$K_\mathrm{B}T$), $E_\mathrm{a}$ is the activation energy) with $E_\mathrm{a} \approx$ 80 meV, which is very similar to the value observed in the bulk ~\cite{Zhou:2005p7395,Katsufuji:1997p10145}. Further transport analysis strongly suggests that the antiferromagnetic transition temperature ($T_N \sim$ 120 K) is very similar to the bulk PTO [see SM~\cite{SM}]. We performed the reciprocal space mapping and the crystal truncation rod measurements at the 33ID beamline of the Advanced Photon Source, USA. XLD measurements were carried out at the beamline 4.0.2 of the Advanced Light Source, USA using  total electron yield mode (probing depth at least 5 nm for Ti $L_{3,2}$ edges). To minimize surface oxidation, a 10 uc PTO film, capped by 2.5 nm amorphous AlO$_x$ has been  used for these synchrotron measurements.  Further, we performed first-principles density functional theory (DFT)\cite{DFT} calculations (for details see Ref.~\cite{dft_extra}).

	Our reciprocal space mapping using synchrotron X-ray demonstrates that the PTO film is coherently strained with the underlying substrates [Fig ~\ref{Fig1}(a) and (b)]. The key ingredient behind the proposed mechanism of FOO in antiferromagnetic $RE$TiO$_3$ is the presence of antiparallel displacement of  $RE$-sites due to GdFeO$_3$-type distortion~\cite{Mochizuki:2003p167203} [also see SM~\cite{SM}].
	To determine the $RE$-site displacements and octahedral rotational pattern (ORP), half order crystal truncation rods have been examined by synchrotron XRD.  The presence of the (0 1/2 2)$_\mathrm{pc}$ and the absence of the (1/2 0 2)$_\mathrm{pc}$ Bragg peak  for both NGO substrate and the PTO film [Fig. ~\ref{Fig1}(c)] exclusively demonstrate the anti-parallel displacement of $RE$ sites within both film and substrate with same in-plane orientation. Diffraction measurements around certain half order Bragg reflections [Fig ~\ref{Fig1}(d)] have found  $a^-b^+c^-$ ORP for our PTO film, similar to the bulk~\cite{COBRA}.

	\begin{figure}
		\vspace{-0pt}
		\includegraphics[width=0.8\textwidth] {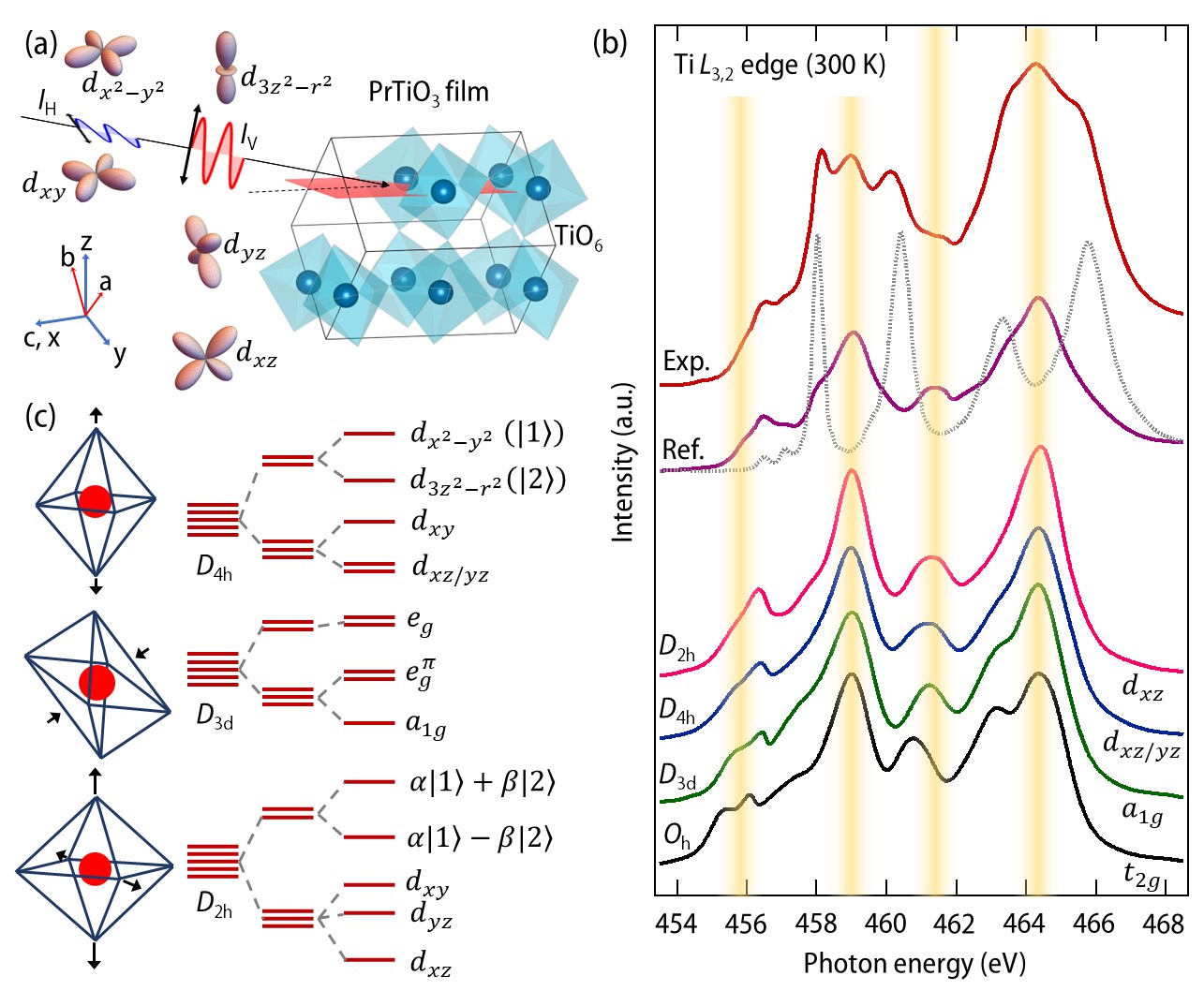}
		\caption{ \label{Fig2} {(a) Schematic of the XLD experiment in grazing incidence geometry (with incident angle of 20$^{\circ}$  from the (110)$_\mathrm{or}$ plane, shown in red). The crystallographic axes are denoted as (a, b, c) whereas the experiment and the simulations are performed in pseudocubic (x, y, z) coordinate.  (b) Experimentally observed XAS data at Ti $L_{3,2}$ edge at 300K (red). For comparison, the Ti$^{3+}$ (purple) and Ti$^{4+}$ (gray) reference spectra, normalized to equal integral spectral weight, are taken from ~\cite{Scheiderer:2018p1706708}. Considering the experimental configuration, Ti-$L_{3,2} $XAS is simulated  for orthorhombic ($D_{2h}$), tetragonal ($D_{4h}$), trigonal ($D_{3d}$), octahedral ($O_h$) symmetry.    The corresponding resulting splitting of the $d$ levels have been shown in panel (c).
		}}
	\end{figure}
	
	To directly probe the orbital structure, we have chosen XLD spectroscopy. In this technique, X-ray absorption spectra are recorded with linearly polarized X-rays having the electric field vector parallel and perpendicular to the sample surface~\cite{Chakhalian:2007p1114,Pesquera:2012p1189,Salluzzo:2009p166804,Salluzzo:2013p087204,Salluzzo:2009p166804,Cao:2016p16009,Lee:2013p703,Stohr:2006magnetism}. Due to the strong dependence of the absorption process on the angle between photon polarization vector and crystallographic axis,   Ti $L_{3,2}$ (2$p\rightarrow 3d$) XAS for $E||xy$ ($I_H$) and $E||z$ ($I_V$) [Fig.~\ref{Fig2}(a) for experimental geometry] probes the unoccupied in-plane [$d_{xy}$, $d_{x^2-y^2}$] and out-of-plane [$d_{xz}$, $d_{yz}$, $d_{3z^2-r^2}$] orbitals, respectively~\cite{Stohr:2006magnetism}. Thus, the sign of XLD (=$I_V$-$I_H$) can disclose even small energy splitting between sub-bands and their orbital characters~\cite{Salluzzo:2009p166804,Salluzzo:2013p087204,Lee:2013p703,Cao:2016p16009, Mandal:2021pL060504}, and XLD technique is used very often to probe orbital ordering~\cite{Huang:2004p087202, Iga:2004p257207, Chen:2010p201102}. We have measured XLD at 30 K and 300 K to probe the orbital state in the antiferromagnetic and paramagnetic phases, respectively~\cite{Haverkort:2005p056401,Komarek:2007p224402,Cheng:2008p087205,Zhang:2020p035113}.

	The average Ti XAS [($I_V$+$I_H$)/2], which probes the oxidation state, has been shown in Fig. ~\ref{Fig2}(b). Comparing our experimental data at 300 K with the reference spectra for Ti$^{3+}$ and Ti$^{4+}$~\cite{Scheiderer:2018p1706708}, we observe XAS features of the desired Ti$^{3+}$ together with unwanted Ti$^{4+}$ in our PTO film. Such Ti$^{4+}$ appears due to surface oxidation and is very common in $RE$TiO$_3$ thin films~\cite{Scheiderer:2018p1706708,Aeschlimann:2018p1707489}. Using these Ti$^{3+}$ and Ti$^{4+}$ reference XAS spectra with equal integrated spectral weight, we find that 75-80\% Ti are with +3 oxidation state [analysis shown in  SM~\cite{SM}]. The absence of any impurity phase in the (0 0 $L$)$_\mathrm{pc}$ scan of XRD [see SM~\cite{SM}]  and the confirmation of bulk PTO-like electrical transport behavior and orthorhombic symmetry of the ultrathin PTO film strongly suggest that the excess interstitial oxygen atoms are distributed randomly in the film without affecting its electronic and structural behaviors. We have focused on analyzing XAS and XLD features related to the Ti$^{3+}$,  which are highlighted by yellow shade in Fig.~\ref{Fig2}(b).

	The XAS line shape strongly depends on the local structural symmetry of the TiO$_6$ unit. In the absence of any octahedral tilt and rotation,  the wave function of the lowest occupied state ($a_{1g}$ orbital) for $D_{3d}$ symmetry  (Fig. ~\ref{Fig2}(c)) is 1/$\sqrt 3$ ($d_{xy}$+$d_{yz}$+$d_{xz}$)~\cite{Mochizuki:2003p167203}. The low temperature NMR spectra were explained assuming the existence of such orbitals ~\cite{Kiyama:2003p167202}. However,  TiO$_6$ octahedron does not follow the $D_{3d}$ symmetry~\cite{Eitel:1986p95,Komarek:2007p224402}, and the symmetry should rather be $D_{2h}$ ~\cite{Solovyev:2004p134403}.
	Several theoretical studies also predicted that the occupied orbital of Ti has dominant ($d_{yz}$+$d_{xz}$) orbital character  in the presence of GdFeO$_3$-type distortions~\cite{Pavarini:2004p176403,Varignon:2017p235106}.
	The presence of epitaxial strain in thin film samples will add further complexity to the problem. For example, in the present case, the compressive strain should induce a tetragonal distortion ($D_{4h}$ symmetry). We have simulated $L_{3,2}$-edge XAS spectra of Ti$^{3+}$ using ligand field multiplet theory based Quanty program for $O_{h}$, $D_{3d}$, $D_{2h}$ and $D_{4h}$ symmetry~\cite{Haverkort:2012p165113}. The value of all parameters used in the simulations have been listed in Ref. ~\cite{para}.

	The spectrum, especially the features within 454-456 eV, simulated with $O_h$ symmetry doesn't agree  with the observed Ti$^{3+}$ spectra~\cite{Haverkort:2005p056401}, indicating a lifting of degeneracy for the $t_{2g}$  states (Fig. ~\ref{Fig2}(b)). Our simulated spectra for $D_{3d}$ with occupied $a_{1g}$ state and $D_{4h}$ with occupied $d_{xy}$/$d_{yz}$ state show a good agreement with the reference spectra (Fig.~\ref{Fig2}(b)). The simulated spectrum for $D_{2h}$ with occupied $d_{xz}$ state also matches very well with the reference one. However, the simulated spectrum of $D_{2h}$ symmetry with either $d_{xy}$ or $d_{yz}$ as the lowest orbital is very different [see SM~\cite{SM}]. We further note that a gap ($\Delta_\mathrm{gap}$) of 0.35-0.50 eV between the lowest occupied and the next unoccupied state was necessary for each $D_{3d}$, $D_{4h}$ and $D_{2h}$ symmetry  to match the spectra. These results demonstrate that while isotropic XAS can definitively conclude the existence of non-cubic crystal field, it fails to determine  the symmetry of the occupied orbital.

	\begin{figure}
		\vspace{-0pt}
		\includegraphics[width=0.6\textwidth] {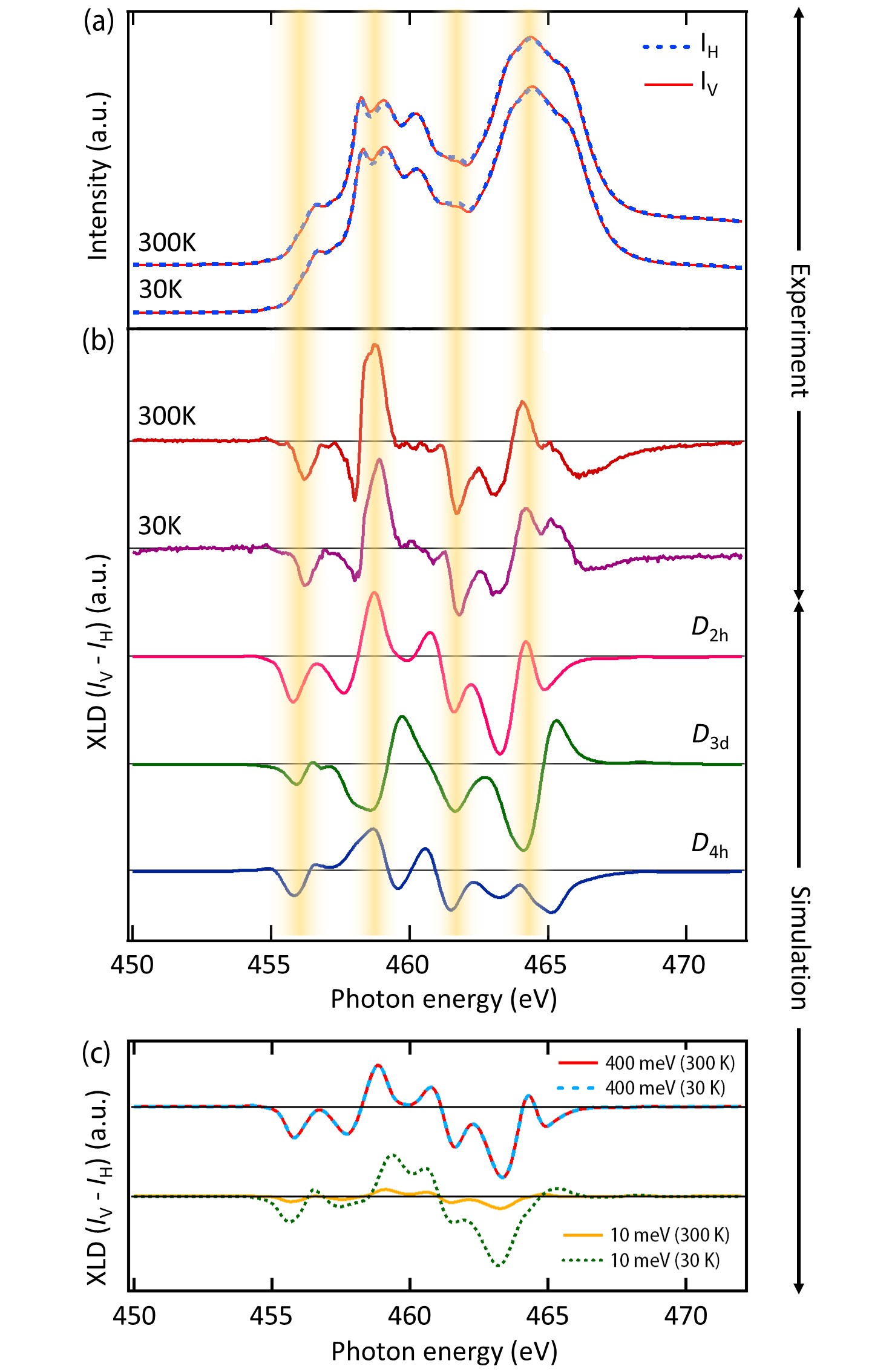}
		\caption{ \label{Fig3} {(a) Ti $L_{3,2}$ edge XAS spectra recorded in two different polarizations (I$_V$ and I$_H$) at 300 K and at 30 K. (b) shows the experimental XLD (I$_V$ - I$_H$) data at 300 K and 30 K and simulated XLD data in experimental geometry for different symmetries of the TiO$_6$ octahedron. We  also note that the sign of simulated XLD line shape remains unaffected in presence of the octahedral tilting [see SM~\cite{SM}]. (c) XLD spectra in $D_{2h}$ symmetry is simulated for $\Delta_\mathrm{gap}$ = 400 meV (top) and 10 meV (bottom) at 300 K and 30 K. } }
	\end{figure}

	To find out the orbital state, we have focused on polarization dependent absorption spectra and XLD ([Fig.~\ref{Fig3}(a) and (b)) at the paramagnetic phase, where XLD is contributed  by anisotropic  charge distribution only~\cite{Stohr:2006magnetism}. In order to understand the origin of finite XLD feature, we have simulated  XLD spectra for the three different non-cubic crystal fields considering the experimental geometry [denoted by xyz coordinate system in Fig.~\ref{Fig2}(a)]. In the $D_{4h}$ symmetry the line shape of the simulated XLD spectra is markedly different from the experimental data. In $D_{3d}$ symmetry with $a_{1g}$ as the lowest orbital, large deviations are observed around  459 eV and 464 eV.
	As evident, the spectra simulated in the $D_{2h}$ symmetry with $d_{xz}$ as the lowest orbital shows best agreement with our experimental observation. Moreover, simulated XLD of $D_{2h}$ symmetry for alternative cases with $d_{yz}$ and $d_{xy}$ as the lowest state do not match with our experimental observation (see SM~\cite{SM}). This establishes that the electron occupies predominantly  $d_{xz}$ orbital  in $D_{2h}$ symmetry.
	To further testify whether any arbitrary value of $\Delta_\mathrm{gap}$ within $D_{2h}$ symmetry can result orbital polarization, we have simulated a series of XLD spectra [shown in SM~\cite{SM}] and find that spectra with $\Delta_\mathrm{gap}\sim$ 300-500  meV only match with our experimental observation. This implies that the electron occupies a  non-degenerate orbital, that is energetically well separated from remaining ones. This energetics should lock the orbitals spatially, resulting in a FOO rather than OL, in the paramagnetic phase. Such large orbital splitting results quenching of the orbital moment~\cite{Haverkort:2005p056401}, which would lead to temperature insensitivity of XAS and XLD line shape. Indeed, we observe the same in our experiments (Fig.~\ref{Fig3}(a) and (b)), further corroborated by temperature dependent (30 K and 300 K) XLD simulation with large gap (Fig.~\ref{Fig3}(c), $\Delta_\mathrm{gap} =$ 400 meV). However, the XLD spectra is strongly temperature dependent for smaller gap ($\Delta_\mathrm{gap} = $ 10 meV) [also see SM~\cite{SM}]. Thus, we establish that in presence of large orthorhombic distortion, the same orbital ordering pattern is also preserved in the antiferromagnetic phase.

	\begin{figure}
		\centering
		\includegraphics[width=0.6\textwidth]{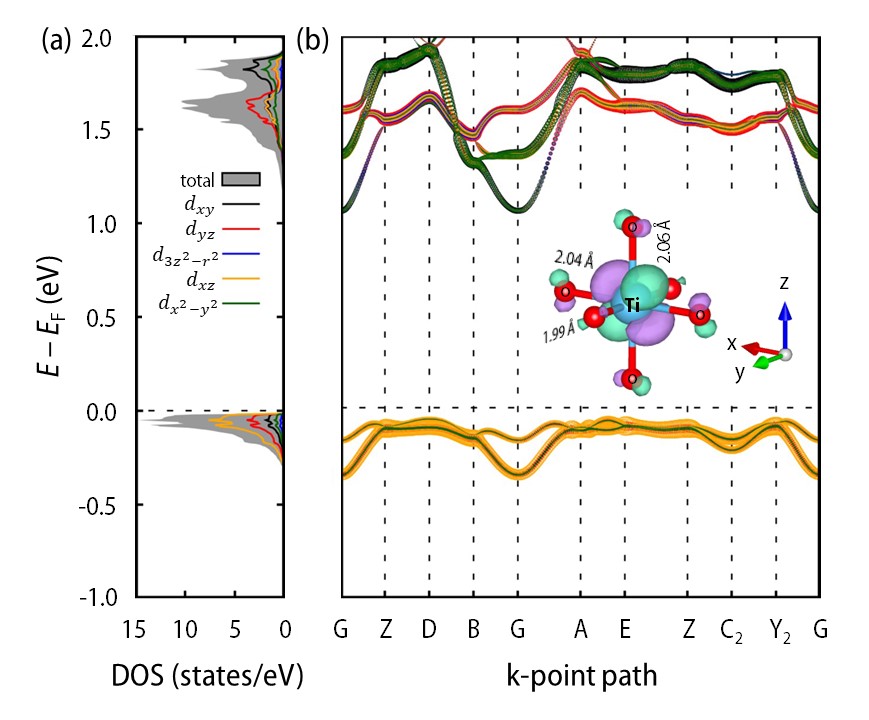}
		\caption{ Calculated DOS (a) and electronic band structure (b) for majority spin channel. $lm$-decomposed data are
		plotted in the TiO$_6$ local basis. Black, red, blue, yellow, and green curves (dots) represent $d_{xy}$, $d_{yz}$, $d_{3z^2-r^2}$, $d_{xz}$, $d_{x^2-y^2}$, respectively. Fermi level is set to zero. The inset of (b) shows the MLWFs obtained on each TiO$_6$ octahedron.  The constant value surfaces with lobes of different signs are colored as purple and cyan.}
		\label{fig:dft}
	\end{figure}

Our XLD experiments on thin film of PrTiO$_3$, in conjunction with cluster calculation, have demonstrated the existence of ferro orbital ordering. To testify whether epitaxial strain plays any role in stabilizing this FOO phase, we performed first-principle studies for PrTiO$_3$ with (i) bulk lattice constants and (ii)  our thin film lattice parameters.  A TiO$_6$ octahedron from the relaxed structure for the second case is shown in the inset of Fig. \ref{fig:dft}.  According to the bond length differences, the six Ti-O bonds around the center Ti atom can be classified into three categories, and hence, we defined the local basis of all the TiO$_6$ octahedra in the following way: $z$- and $y$-axes are along the directions with the longest and the shortest Ti-O bond length, respectively (2.06 and 1.99 \AA) whereas the $x$-axis is along the other direction (with the bond length equals to 2.04 \AA), which give each octahedron a  local $D_{2h}$ symmetry. As shown in the electronic density of states  (Fig.~\ref{fig:dft}(a)) and band structure (Fig.~\ref{fig:dft}(b)) plot, the occupied states of the Ti atom have mostly $d_{xz}/d_{yz}$ orbitals with a dominant contribution from the $d_{xz}$ orbitals while the $d_{3z^2-r^2}$ and $d_{x^2-y^2}$ orbitals are almost degenerate with a nearly zero occupation.  The two bands just below the Fermi level are represented by the maximally-localized Wannier functions (MLWFs). As it can be seen in the inset of Fig.\ref{fig:dft} (b), within the proper TiO$_6$ local basis, the most occupied natural orbitals have a perfect $d_{xz}$ shape. Therefore, these results are in excellent agreement with the experimental finding of  $d_{xz}$ character of the occupied non-degenerate orbital. We have further plotted charge isosurfaces corresponding to the two occupied bands located just below the Fermi energy, which also demonstrates the existence of FOO [see SM~\cite{SM}].
The calculation of bulk PTO yields very similar DOS features [shown in SM~\cite{SM}], demonstrating that the observation of FOO is not linked to the underlying epitaxial strain. Additionally, our first-principle simulations show that varying magnetic states leads to minor modifications of the electronic structure, the $d_{xz}$ character of the occupied orbital remains unaffected by such changes, further demonstrating the dominant contribution of the orthorhombic distortion [see SM~\cite{SM}].

	\section*{Conclusions}

	To conclude, our present work conclusively demonstrates that the $D_{2h}$ crystal field splitting of PrTiO$_3$ is strong enough to have a nondegenerate occupied orbital  in both paramagnetic and antiferromagnetic phases, which is well separated ($\Delta_\mathrm{gap} \approx$ 400 meV) from the remaining orbital manifold. Most importantly, this energy gap is larger than the energy scales of the orbital fluctuations coupled to the lattice vibrations~\cite{Reedyk:1997p1442}, superexchange interaction driven spin-orbital resonance, estimated from orbital liquid model~\cite{Khaliullin:2000p3950}, as well as the  thermal energy scale, leading to orbital ordering. Therefore, we conclusively demonstrate the existence of orbital ordered phase both in the paramagnetic and antiferromagnetic phases of PrTiO$_3$, which is primarily contributed by the orthorhombic distortion, settling a long standing debate about the spin-orbital physics of antiferromagnetic $RE$TiO$_3$.

\section*{Acknowledgements}	
	The authors acknowledge AFM, XRD, wire bonding facilities of the Department of Physics, IISc Bangalore. SM acknowledges DST Nanomission grant  (DST/NM/NS/2018/246) and SERB Core Research grant (CRG/2022/001906) for  financial support.  This
	research used resources of the Advanced Photon Source, a U.S.
	Department of Energy Office of Science User Facility operated
	by Argonne National Laboratory under Contract No. DE-AC02-
	06CH11357. This research used resources of the Advanced Light
	Source, which is a Department of Energy Office of Science User
	Facility under Contract No. DE-AC02-05CH11231. The computations were enabled in project SNIC 2021/3-38 by resources provided to BS by the Swedish National Infrastructure for Computing (SNIC) at NSC, PDC and HPC2N partially funded by the Swedish Research Council through grant agreement no. 2018-05973. BS acknowledges allocation of supercomputing hours by PRACE DECI-17 project `Q2Dtopomat' in Eagle supercompter in Poland and EuroHPC resources in Karolina supercomputer in Czech Republic. BS and SM acknowledge Indo-Swedish Joint Network Grant 2018 provided by Swedish Research Council (grant no. 2018-07082). BS also acknowledges financial support from Swedish Research Council (grant no. 2022-04309).

%


\newpage
\setcounter{figure}{0}
\renewcommand{\thefigure}{S\arabic{figure}}

\makebox[\textwidth]{\bf \Large Supplementary Information}

\hspace{1cm}

	\begin{flushleft}
	\textbf{S1. Definition of unit cell in pseudocubic notation}
\end{flushleft}

\begin{figure}[h]
	\vspace{-0pt}
	\includegraphics[width=1\textwidth] {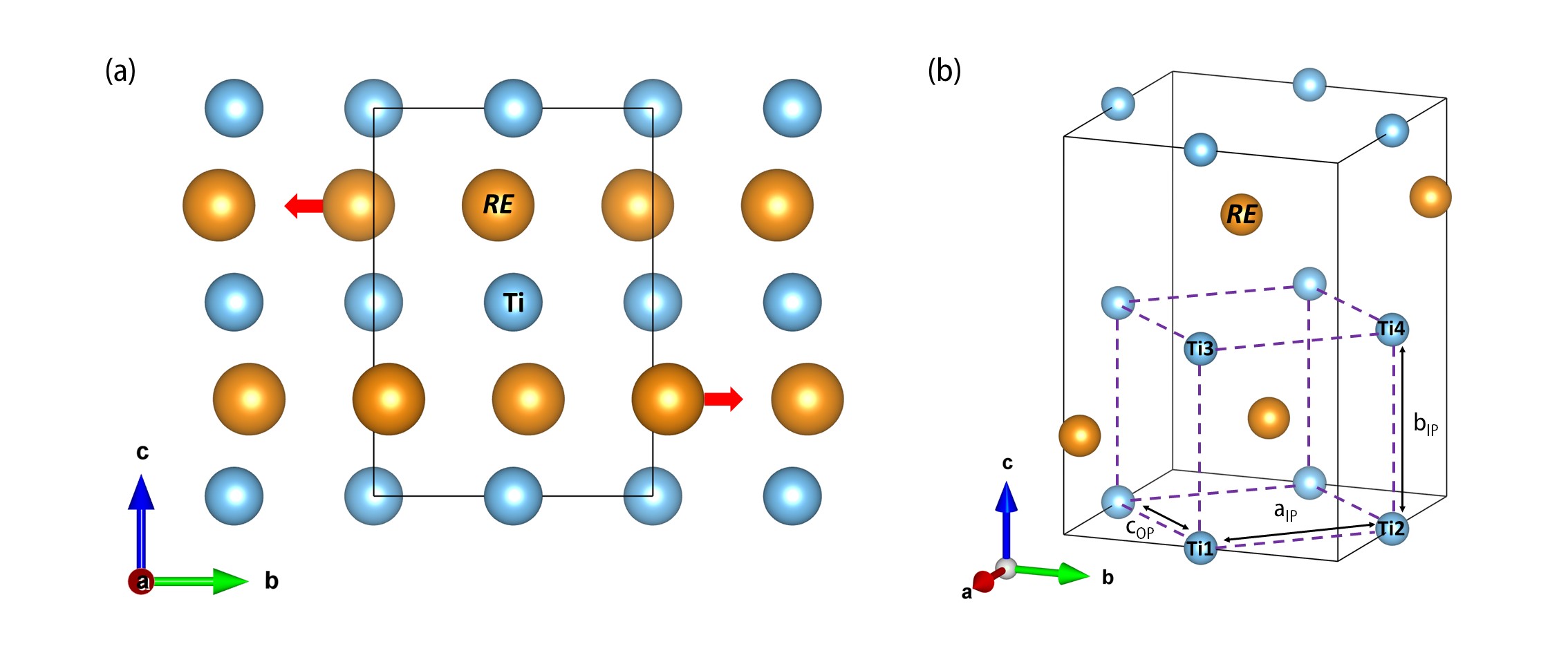}
	\caption{ \textbf{Antiparallel displacement and pseudocubic unit cell of $RE$TiO$_3$.}  \label{pseudocubic} { (a) The antiparallel displacement of the $RE$ ions  is shown by red arrows. (b) The orthorhombic unit call (solid lines) and the pseudocubic unit cell (dashed lines) denoted by the lattice parameters ($a$, $b$, $c$) and ($a\mathrm{_{IP}}$, $b\mathrm{_{IP}}$, $c\mathrm{_{OP}}$) respectively. Ti1, Ti2, Ti3 and Ti4 corresponds to the four inequivalent Ti sites.}}
\end{figure}

Bulk PrTiO$_3$ (PTO) and NdGaO$_3$ (NGO), both have orthorhombic $Pbnm$ symmetry. In such crystal structures, the $RE$ sites (Pr, Nd) are displaced antiparallelly along the crystallographic $b$ direction [Fig.~\ref{pseudocubic} (a)]. The PTO film is grown on NGO (1 1 0) (orthorhombic setting) substrate.  Orthorhombic and pseudocubic unit cell  has been denoted in Fig.~\ref{pseudocubic} (b) and the bulk lattice parameters of PTO and NGO in both orthorhombic ($a$, $b$, $c$) and pseudocubic setting ($a\mathrm{_{IP}}$, $b\mathrm{_{IP}}$, $c\mathrm{_{OP}}$) are listed in Table 1. Here, the pseudocubic notation is defined as, $a\mathrm{_{IP}}$ = $c\mathrm{_{OP}}$ = $\frac{\sqrt{a^2 + b^2}}{2}$ , and $b\mathrm{_{IP}}$ = $\frac{c}{2}$. The values of the lattice parameters of NGO an PTO are taken from Ref. ~\cite{Vasylechko:2000p46} and ~\cite{Zhou:2005p7395}, respectively.

\begin{table}[h]
	\begin{center}
		\caption{Bulk lattice parameters of NdGaO$_3$ and PrTiO$_3$.}
		\begin{tabular}{c c c c c c} 
			\hline
			& $a$ (\AA) & $b$ (\AA)  & $c$ (\AA) & $a\mathrm{_{IP}}$ = $c\mathrm{_{OP}}$ (\AA) & $b\mathrm{_{IP}}$ (\AA) \\
			\hline
			\hline
			NdGaO$_3$ & 5.428 & 5.498 & 7.708 & 3.863 & 3.854 \\
			PrTiO$_3$ & 5.555 & 5.615 & 7.821 & 3.949 & 3.911 \\
			\hline
			\label{t1}
		\end{tabular}
	\end{center}
\end{table}

\clearpage

\begin{flushleft}
	\textbf{S2. Growth and characterization of PrTiO$_3$ thin film}
\end{flushleft}

\begin{figure}[h]
	\vspace{-0pt}
	\includegraphics[width=0.85\textwidth] {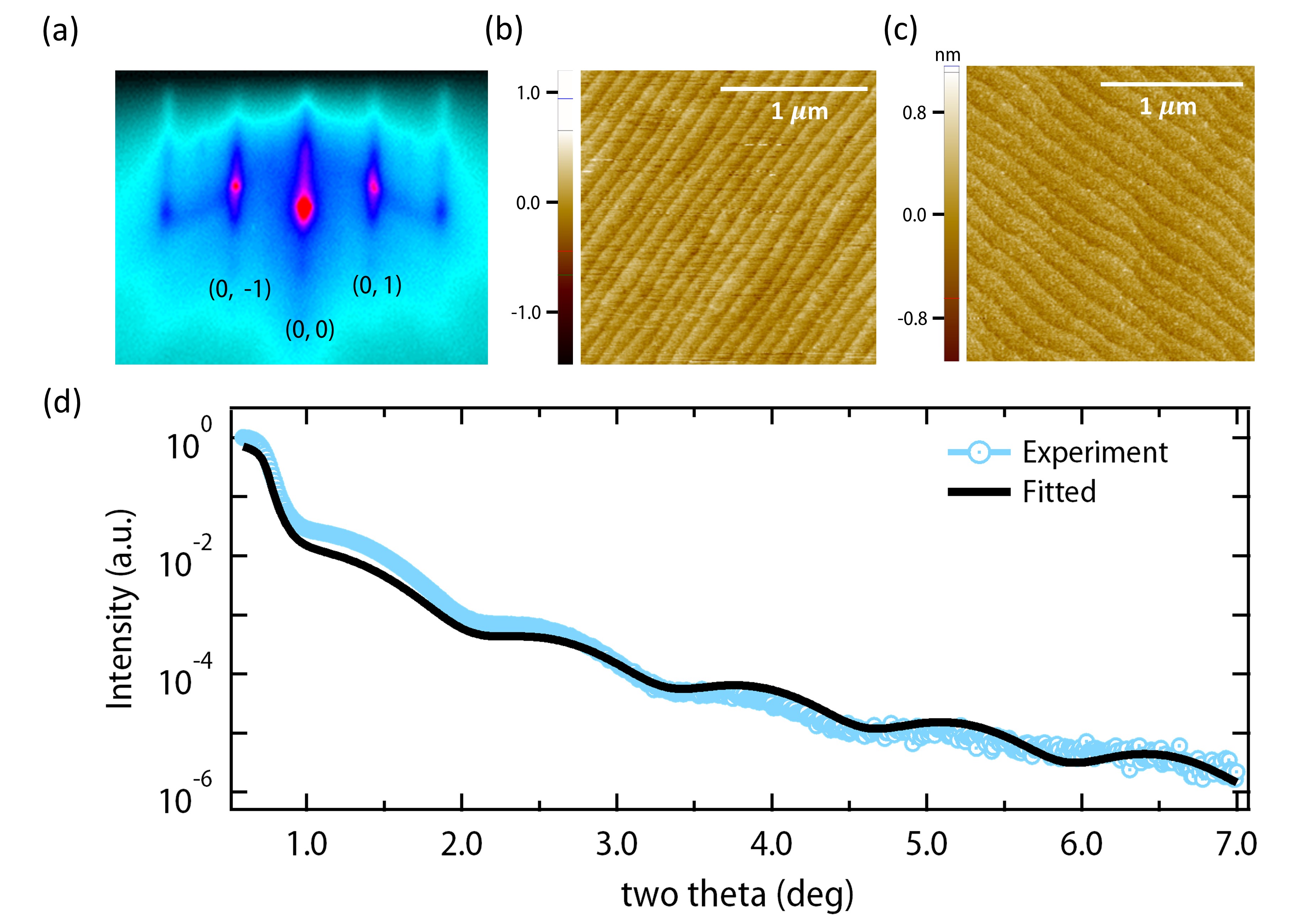}
	\caption{ \textbf{Surface and interface characterization of the PTO film.} \label{growth} {(a) Reflection high energy electron diffraction (RHEED) image of the PTO thin film.  (b) and (c) are atomic force microscopy image of the surface of the treated substrate and the film. (d) The X-ray reflectivity data (XRR) of the PTO on NGO substrate.}}
\end{figure}

PTO films  were grown on single crystalline NdGaO$_3$ substrate using a pulsed laser deposition (PLD) system, connected with a RHEED (reflection high energy electron diffraction)  set up~\cite{Mandal:2021p12968}. For growing the PTO film, a polycrystalline Pr$_2$Ti$_2$O$_7$ target, prepared by solid state synthesis method, was ablated by a KrF excimer laser (248 nm), operating at a repetition rate of 2 Hz. During the growth, the heater was maintained at 850$\mathrm{^o}$C while the chamber pressure was $\approx$ 10$^{-6}$ Torr. The streak pattern of specular (00) and  off-specular (0, $\pm$1) reflection in RHEED image confirm smooth surface morphology (Fig.~\ref{growth} (a)).
To minimize unintentional doping of excess oxygen, observed routinely in  $RE$TiO$_3$ thin films~\cite{Aeschlimann:2018p1707489, Scheiderer:2018p1706708}, a 10 uc  PTO film (thickness $\sim$ 4 nm) used in synchrotron measurement was capped by 2.5 nm amorphous Al$_2$O$_3$ layers before taking it out  from  the PLD chamber. A commercially available single crystalline Al$_2$O$_3$ target was used for growing the amorphous capping layer at room temperature.

The surface of the as-received NGO substrate is mixed terminated. In order to get a single terminated surface, we have annealed the substrate at 900$^{\circ}$C, prior to the growth. The surface topography of such a substrate, imaged with a Park systems atomic force microscope, shows step-like features thus  (Fig.~\ref{growth}(b)) confirming the  realization of single terminated substrate. The step features remain unchanged after the completion of the growth (Fig.~\ref{growth}(c)), signifying desired layer-by-layer growth of the PTO film.

The film was also characterized by X-ray reflectivity (XRR) measurement using Rigaku Smartlab diffractometer. By fitting the XRR data (Fig.~\ref{growth}(d)) using GenX program~\cite{Bjorck:2007p1174}, we obtain the film/substrate interface roughness around 2.1 \AA\ . Thus our AFM and XRR measurements testify excellent morphological quality of the film. 

\clearpage

\begin{flushleft}
	\textbf{S3. DC electrical transport behavior}
\end{flushleft}

\begin{figure}[h]
	\vspace{-0pt}
	\includegraphics[width=0.98\textwidth] {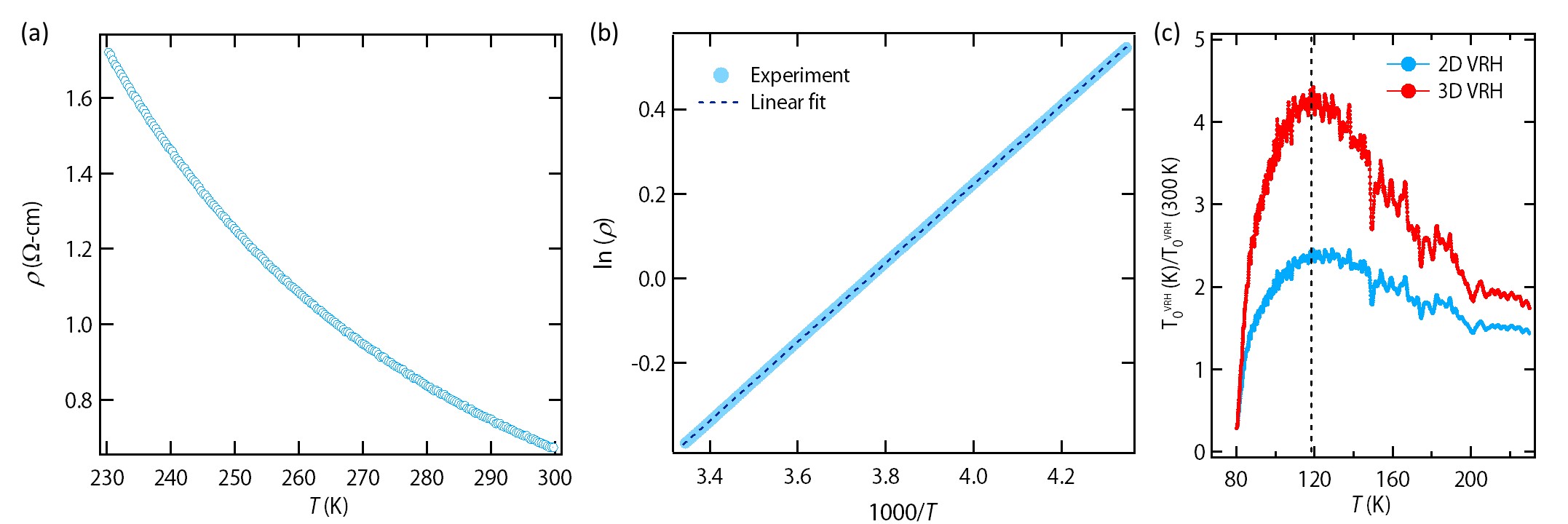}
	\caption{ \textbf{Antiferromagnetic Mott insulating behavior of the PTO film.}  \label{rt} {(a) DC electrical transport behavior of the PTO film. (b) The Arrhenius plot (solid line) and it's linear fitting (dashed line) reveals that the activation energy gap is $\approx$ 80 meV. The data in both (a) and (b) are measured in the four-probe Van der Pauw geometry (c) Temperature dependence of the normalized characteristic temperature scales obtained from the two probe resistance vs temperature data using the Eq. ~\ref{eq3}. }}
	
	\begin{equation} \label{eq3}  
		T_0^{VRH} = \left[\frac{d(lnR)}{d(T^{-1/(d+1)})}\right]^{(d+1)}
	\end{equation}

\end{figure}

DC electrical transport measurements (Fig.~\ref{rt} (a) and (b)) were performed in a four-probe Van der Pauw method using a Keithley 2450 Source measurement unit and a ARS (Advanced Research Systems) close cycle cryostat. In the high temperature regime (230 K - 300 K), the transport behavior shows an activated behavior (Eq.~\ref{eq1}). A deviation from such behavior is observed at lower temperature where the resistance starts increasing rapidly and crosses the measurement limit of the instrument. Thus, for the low temperature measurement, we have used a Keithley 2002 Multimeter to measure the two probe resistance (limit $\approx$ 1 G$\Omega$). To understand the transport behavior, we have assumed the 2D and 3D variable range hopping (VRH) models (Eq.~\ref{eq2}, d = 2 and 3 corresponds to 2D and 3D respectively).

\begin{equation} \label{eq1}
	R = R_0exp(E_g/k_BT)
\end{equation}

\begin{equation} \label{eq2}
	R = R_0exp\left(\left(\frac{T_0^{VRH}}{T}\right)^{1/(d+1)}\right)
\end{equation}


In Fig. ~\ref{rt} (c), the characteristic temperature (T$_0^{VRH}$, obtained from Eq.~\ref{eq3}), normalized at 300 K, is plotted against the temperature. In case of pure VRH mechanism, T$_0^{VRH}$ is expected to be independent of temperature. The deviation from the such behavior and sharp variation near the antiferromagnetic transition temperature (T$_N$) of bulk PTO ($\approx$ 120 K)~\cite{Zhou:2005p7395} suggests that the spin-dependent scattering contributes to transport behavior, also observed for LaTiO$_3$~\cite{Hemberger:2003p066403}. Thus, the results indicate that T$_N$ of PTO thin film is $\approx$ 120 K, close to that of bulk PTO.

\clearpage

\begin{flushleft}
	\textbf{S4. Synchrotron based X-ray diffraction and fitting}
\end{flushleft}

The structural quality of the thin film is probed by surface X-ray diffraction measurements (SXRD), where the crystal truncation rods (CTRs) are measured using a six-circle diffractometer with X-ray energy of 15.5 keV ($\lambda$=0.7999 \AA) at beam-line 33-ID-D of the Advanced Photon Source, Argonne National Laboratory, USA. The different specular and off-specular CTRs are scanned along the vertical reciprocal space coordinate $L$ (in the reciprocal lattice unit of pseudocubic NGO) with a maximum value of $L_{max}$ = 5.8; for other reciprocal space coordinates, $H_{max}$ = $K_{max}$ = 2. The 2-dimensional image of the diffraction spot at each step of $L$ is recorded with a pixel array area detector (Dectris PILATUS 100 K). To extract the intensity, we have carefully removed the background and integrated over a small region around the diffraction spot for each image followed by a geometric factor correction of the spectra.

\begin{figure}[h]
	\vspace{-0pt}
	\includegraphics[width=0.81\textwidth] {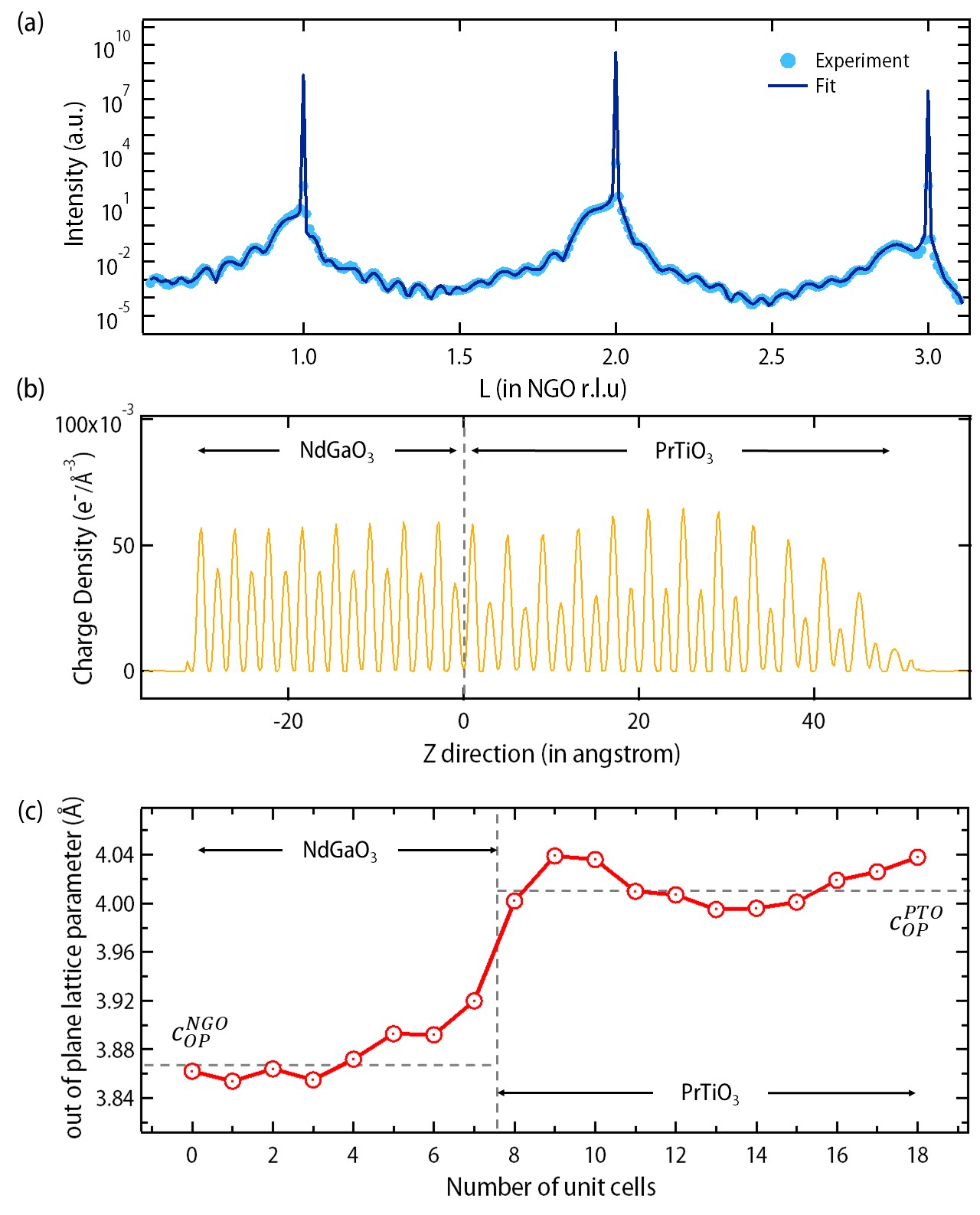}
	\caption{ \textbf{Structural characterization of the film by COBRA.} \label{cobra} {(a) The $00L$ scan (symbol) and the fitted spectra (solid line). (b) The converged electron density obtained from the fitting. (c) The variation of the out-of-plane lattice parameter $c\mathrm{_{OP}}$ across the interface. }}
\end{figure}

Fig.~\ref{cobra}(a) shows the (00$L$) scan, where the intense peaks and their shoulder peaks at lower $L$ values corresponds to the (001), (002) and (003) diffractions from the NGO substrate and the compressively strained PTO film, respectively. The small oscillatory patterns are known as the thickness fringes. Apart from the (001), (002) and (003) peaks from the PTO film, no other peak is observed, indicating the proper perovskite phase of the film. The layer dependent structural information is reconstructed by an iterative phase retrieval method known as coherent Bragg rod analysis (COBRA) which is widely used technique for thin films and other systems~\cite{Fister:2014p021102, Zhou:2010p8103, Kumah:2009p835, Willmott:2007p155502, Disa:2015p026801}. This method is implemented in an in-house written Matlab based code to fit the (00$L$) data. To perform the fitting, a structural model consisting of the substrate and the film is constructed in the pseudocubic convention. An 11 uc PTO film is considered whereas Only 8 uc of the NGO was sufficient for the fitting. In the program, the electron density is allowed to vary until the convergence is achieved. As evident, such model provides excellent agreement with the experiment, further implying the absence of any impurity phase.

Fig.~\ref{cobra}(b) shows the converged charge density across the interface which is denoted by the dashed line. The intense peaks are from the NdO/PrO planes having larger charge density whereas the less intense ones are from the GaO$_2$/TiO$_2$ planes. The position of the planes, after the convergence, is extracted by fitting the peaks with a Gaussian function. Fig.~\ref{cobra}(c) shows the variation of the out of plane lattice parameter ($c_{\mathrm{OP}}$) which is obtained by taking the difference between the position of two consecutive NdO/PrO planes. The dashed line in the substrate region is the bulk lattice parameter of NGO. The mean lattice parameter of the film is shown by dashed line in the PTO region.  As evident, the values of $c_{\mathrm{OP}}$ are very close to the mean value throughout the sample, indicating the coherently strained nature of the film.

\clearpage

\begin{flushleft}
	\textbf{S5. Quantitative estimation of the Ti$^{3+}$/Ti$^{4+}$ content in PTO thin film}
\end{flushleft}

\begin{figure}[h]
	\vspace{-0pt}
	\includegraphics[width=1.0\textwidth] {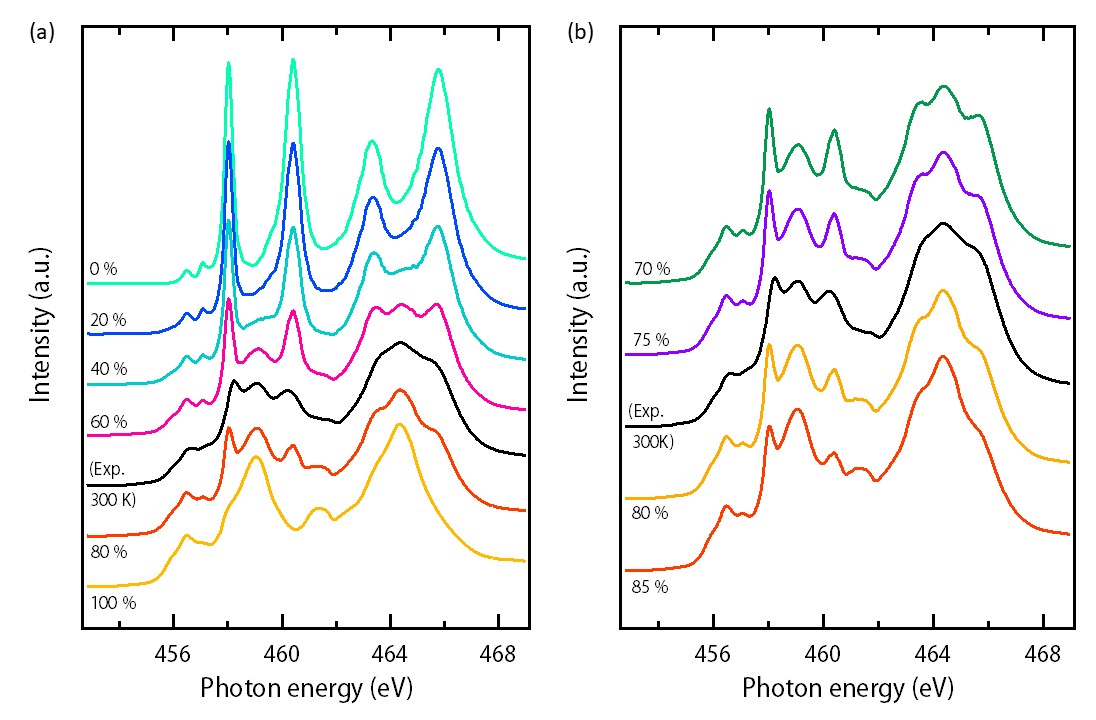}
	\caption{ \textbf{Comparison between experimental and reference XAS spectra.} \label{3+content} Experimentally observed Ti $L_{\mathrm{3,2}}$ XAS data at 300 K (black) compared with different linear combinations of the Ti$^{3+}$ and Ti$^{4+}$ reference spectra (taken from~\cite{ Scheiderer:2018p1706708}) having different percentages of Ti$^{3+}$ content. The percentage of the Ti$^{3+}$ is varied in the interval of 20$\%$ and 5$\%$ in (a) and (b) respectively. }
\end{figure}

The content of Ti$^{3+}$ in our sample is estimated by comparing the experimental XAS data with linear combinations of Ti$^{3+}$ and Ti$^{4+}$ reference spectra which are normalized to equal integrated spectral weight. Fig.~\ref{3+content} (a) clearly demonstrates that approximately 60-80$\%$ Ti$^{3+}$ is present in our samples. The range can be further narrowed down to 75-80$\%$ from Fig.~\ref{3+content} (b). Small mismatch between the experimental data and the reference spectra arises due to the differences in the broadening parameters.

We note that the $RE$TiO$_3$ thin films have a strong tendency to get over-oxidized due to poor stability of Ti$^{3+}$~\cite{Scheiderer:2018p1706708, Aeschlimann:2018p1707489, Aeschlimann:2021p014407, Xu:2016p106803, Cao:2015p112401}. The top surface of the film is usually protected by an inert high band gap insulator (e.g., LaAlO$_3$, AlO$_x$ etc.) to avoid such surface oxidation. However, the presence of a protective layer does not guarantee the presence of 100$\%$ Ti$^{3+}$, which was also demonstrated in ~\cite{Scheiderer:2018p1706708, Aeschlimann:2018p1707489}. We have also shown that we do not find any impurity phase in the synchrotron based XRD. Moreover, the observation of bulk PTO-like activation gap in transport measurement, octahedral roattional pattern etc. imply that these 20-25$\%$ Ti$^{4+}$ are randomly distributed and do not affect the intrinsic electronic and magnetic structure of PTO. 

\clearpage

\begin{flushleft}
	\textbf{S6. Ti 3$d$ level splitting under different symmetries of the crystal field}
\end{flushleft}

\begin{figure}[h]
	\vspace{-0pt}
	\includegraphics[width=1.0\textwidth] {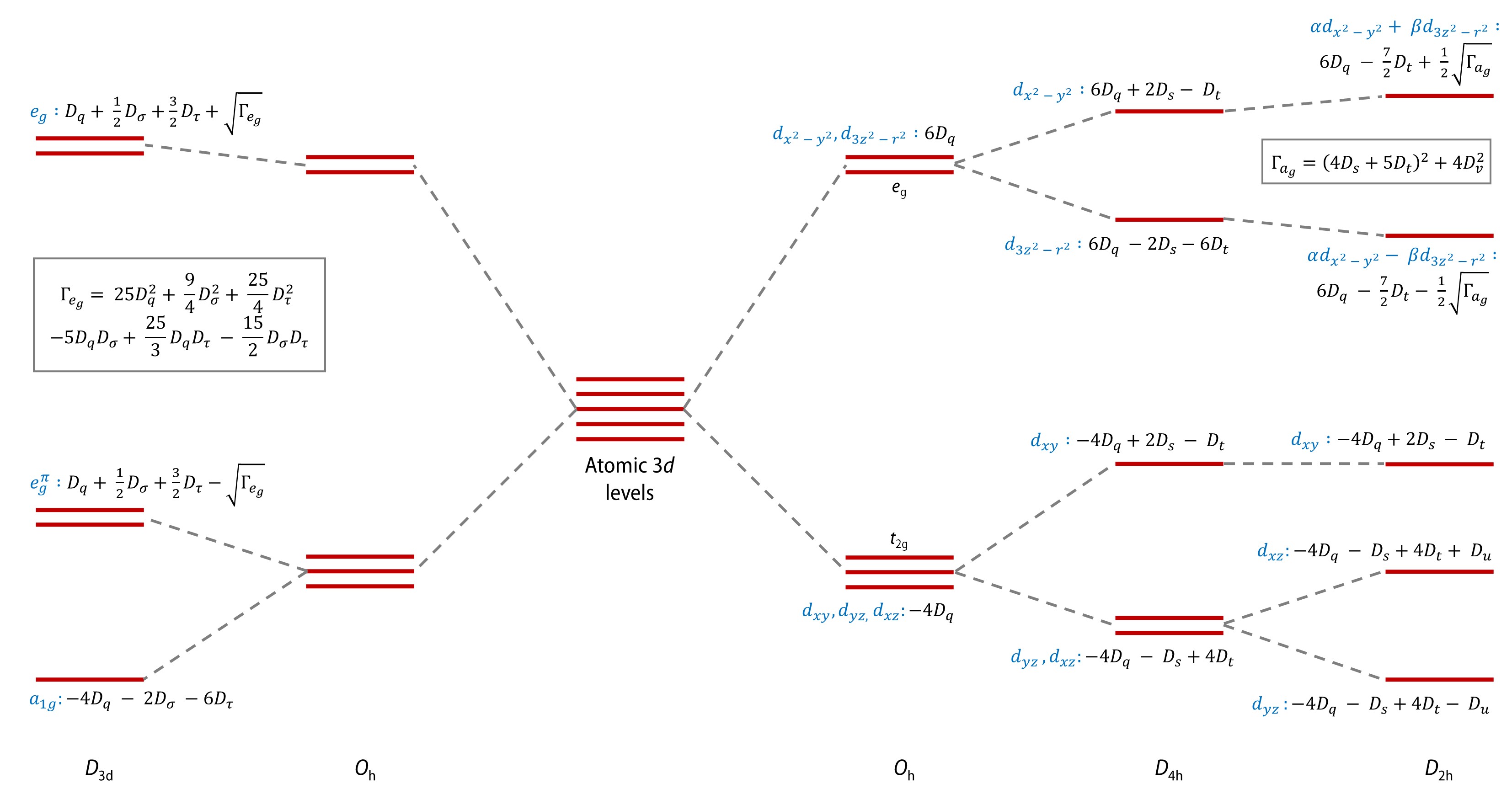}
	\caption{ \textbf{Energy level diagram of the split 3$d$ levels under different crystal fields.} \label{splitting} {The splitting of the degenerate atomic 3$d$ states under crystal field of different symmetries $\textit{viz}$. $O_{h}$, $D_{3d}$, $D_{4h}$ and $D_{2h}$~\cite{Haverkort:2005thesis}.}}
\end{figure}

In a crystalline solid, the degeneracy of the 3$d$ levels in atomic limit is lifted due to the presence of crystal field of the neighboring ions. The energy eigenvalue and the wave function of different split 3$d$ levels (shown in blue in Fig.~\ref{splitting}) are determined by the symmetry of the crystal field. The energy of the different levels are conventionally parameterized by different the parameters: $D_q$, $D_s$, $D_t$, $D_u$, $D_v$, $D_{\sigma}$ and  $D_{\tau}$. The expression for the energy levels and their positions have been shown in Fig.~\ref{splitting}, considering the position of atomic 3$d$ levels at zero energy. For details, we refer to Ref.~\onlinecite{Haverkort:2005thesis}.

\clearpage

\begin{flushleft}
	\textbf{S7. Comparison of the XAS and XLD spectra simulated in $D_{2h}$ symmetry  considering different occupied orbitals}
\end{flushleft}

\begin{figure}[h]
	\vspace{-0pt}
	\includegraphics[width=0.9\textwidth] {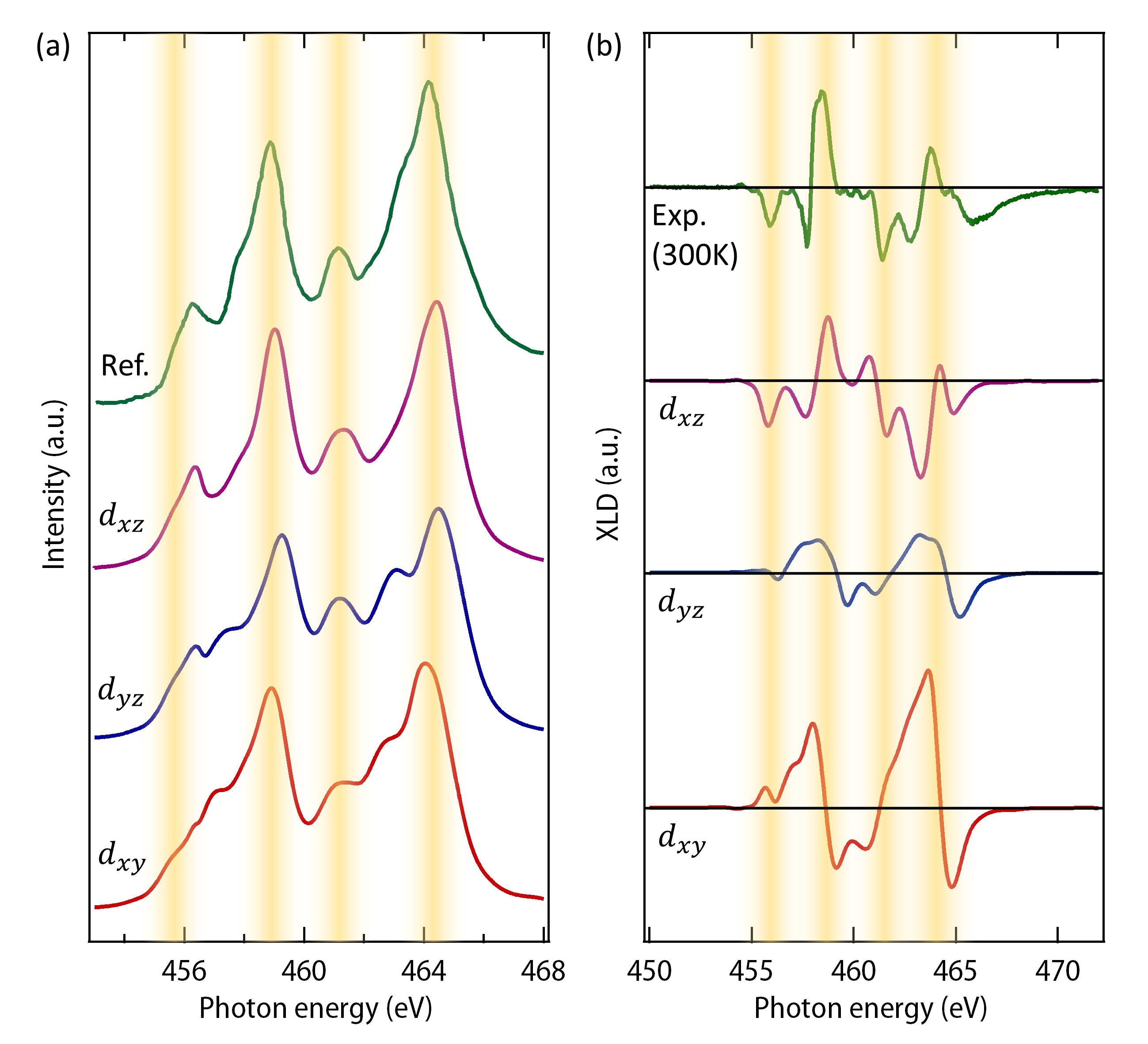}
	\caption{ \textbf{XLD simulation in $D_{2h}$ symmetry with different occupied states.} \label{xz_yx_xy} { The (a) XAS and (b) XLD spectra simulated in the $D_{2h}$ symmetry while the lowest occupied state is varied.  The XAS and XLD spectra are compared with reference Ti$^{3+}$ and experimental XLD spectra, respectively. Yellow shades highlight the Ti$^{3+}$ features.}}
\end{figure}

The XAS and XLD spectra are simulated in the $D_{2h}$ symmetry for three cases where the lowest occupied states are $d_{xz}$, $d_{yz}$ and $d_{xy}$ while the energy difference among them is kept constant. As can be seen in Fig.~\ref{xz_yx_xy}(a), XAS spectra for $d_{xz}$ at the lowest energy matches best with the reference Ti$^{3+}$ spectra. Also, the corresponding XLD spectra (Fig.~\ref{xz_yx_xy}(b)) shows best agreement with our experimental observation.

\clearpage

\begin{flushleft}
	\textbf{S8. Effect of octahedral tilting on XLD}
\end{flushleft}

\begin{figure}[h]
	\vspace{-0pt}
	\includegraphics[width=0.9\textwidth] {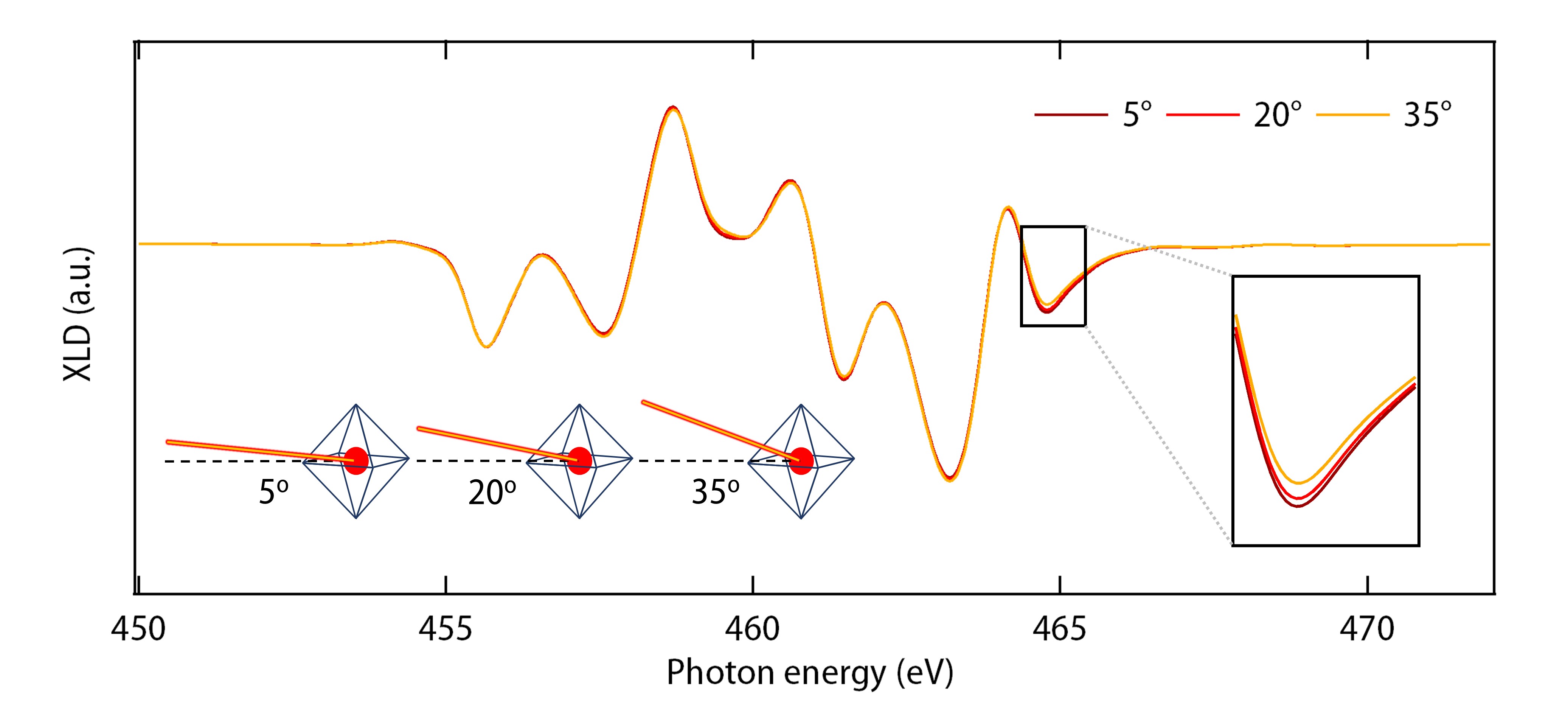}
	\caption{ \textbf{Incident angle dependent XLD simulation in $D_{2h}$ symmetry.} \label{tilt} {The XLD spectra simulated with different angle of incidence of the X-ray in $D_{2h}$ symmetry with $d_{xz}$ as the lowest occupied orbital. }}
\end{figure}

The XAS and XLD simulations based on multiplet ligand field theory are performed on a single TiO$_6$ cluster where a Ti atom is surrounded by six O atoms in different arrangements giving rise to different crystal fields such as $O_h$, $D_{4h}$, $D_{2h}$ and $D_{3d}$. In the absence of octahedral rotation, e.g. SrTiO$_3$, the angle of incidence of X-ray (20$^{\mathrm{o}}$ in our experiment) with respect to the local octahedral coordinate (the three axes along the bond lengths) is same for all the TiO$_6$ octahedra for XAS measurements. However, in orthorhombic crystal having $Pbnm$ symmetry, the octahedra are tilted due to GdFeO$_3$ distortion which makes them inequivelent with respect to each other. Since orthorhombic bulk PrTiO$_3$ has Ti-O-Ti bond angle around 152$^{\mathrm{o}}$, the angle of incidence of X-ray for different TiO$_6$ units are within 20$^{\mathrm{o}}\pm$14$^{\mathrm{o}}$ (approximately). We have simulated  XLD for different angle of incidence (20$^{\mathrm{o}}$, 20$^{\mathrm{o}}\pm$15$^{\mathrm{o}}$), which shows only a small variation ([Fig.~\ref{tilt}]). This signifies that our conclusion from the multiplet calculation of a single TiO$_6$ cluster is also valid for an orthorhombic structure.

\clearpage

\begin{flushleft}
	\textbf{S9. Determination of the strength of the orthorhombic ($D_{\mathrm{2h}}$) distortion }
\end{flushleft}

\begin{figure}[h]
	\vspace{-0pt}
	\includegraphics[width=0.9\textwidth] {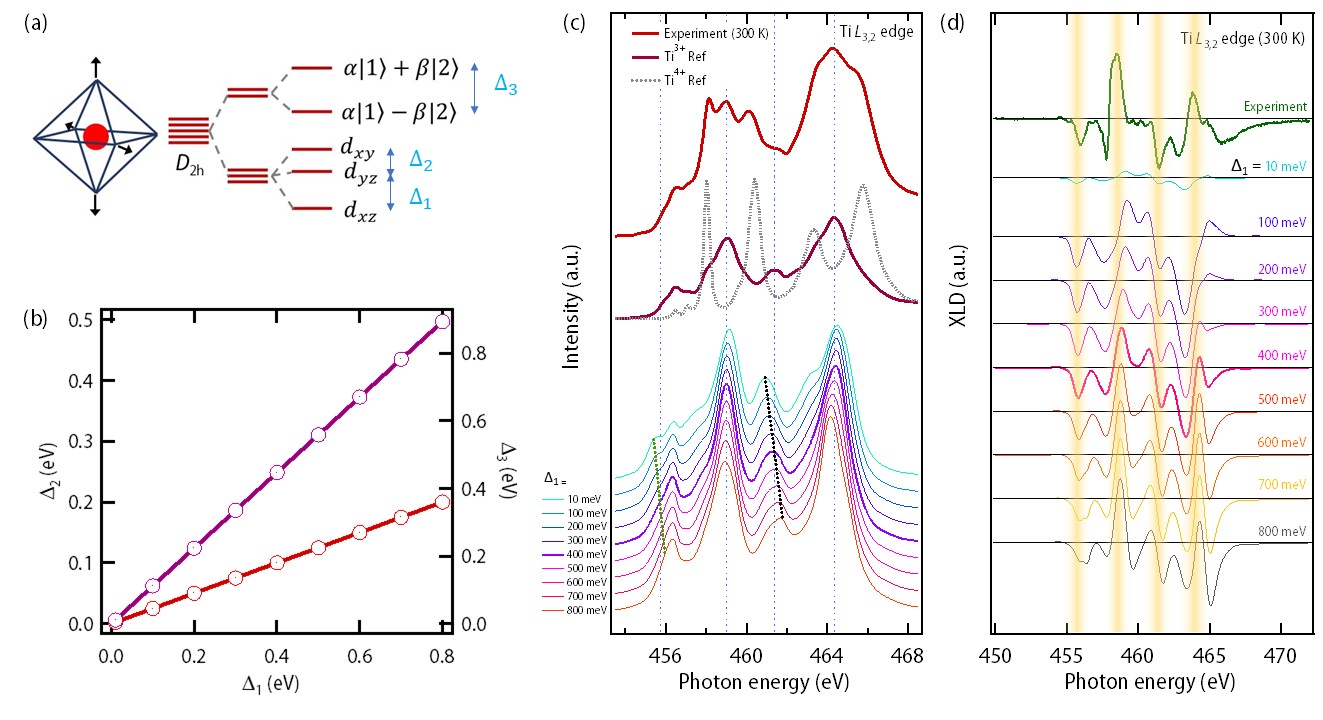}
	\caption{ \textbf{XAS and XLD simulation with varying strength of $D_{\mathrm{2h}}$ distortion.} \label{d2h_vary} { (a) Orthorhombic symmetry ($D_{\mathrm{2h}}$)  of the crystal field and its tuning parameters $\Delta_1$, $\Delta_2$ and $\Delta_3$. (b) The variation of $\Delta_2$ and $\Delta_3$ as a function of  $\Delta_1$. (c) and (d), The XAS and XLD spectra simulated for different strengths of the orthorhomic distortion. }}
\end{figure}

To understand the robustness of orbital ordering against orbital fluctuation, we have performed cluster calculations by varying the strength of the orthorhombic distortion which is parameterized by $\Delta_1$, $\Delta_2$ and $\Delta_3$ (see Fig.~\ref{d2h_vary} (a)). The parameters, $\Delta_2$ and $\Delta_3$, are varied in proportion to $\Delta_1$ (Fig.~\ref{d2h_vary} (b)) that allows the tuning of strength of the orthorhombic distortion  while preserving its symmetry. The XAS and XLD spectra are simulated (Fig.~\ref{d2h_vary} (c) and (d)) by varying the $\Delta_1$ from 10 meV to 800 meV. Fig.~\ref{d2h_vary} (c) clearly shows that spectral features near to 456 eV and 461 eV systematically shifts to higher energy with increasing value of $\Delta_1$. We find that the spectra with $\Delta_1$ = 300-500 eV matches with the experiment. The XLD spectra also shows reasonable agreement with the experiment for the same values of $\Delta_1$. Most importantly, for very small value of $\Delta_1$ (10 meV), the XLD spectra does not match at all, implying the crystal field splitting must be large enough. Such large orthorhombic splitting would would give rise to a nondegenerate lowest occupied orbital that favors the FOO. Further, the value of $\Delta_1\approx$ 400 eV is much larger than the energy scales of orbital fluctuations~\cite{Khaliullin:2000p3950, Reedyk:1997p1442} and thus a robust orbital order should be stabilized.

\clearpage

\begin{flushleft}
	\textbf{S10. Dependence of thermal occupation on orthorhombic distortion }
\end{flushleft}

\begin{figure}[h]
	\vspace{-0pt}
	\includegraphics[width=0.9\textwidth] {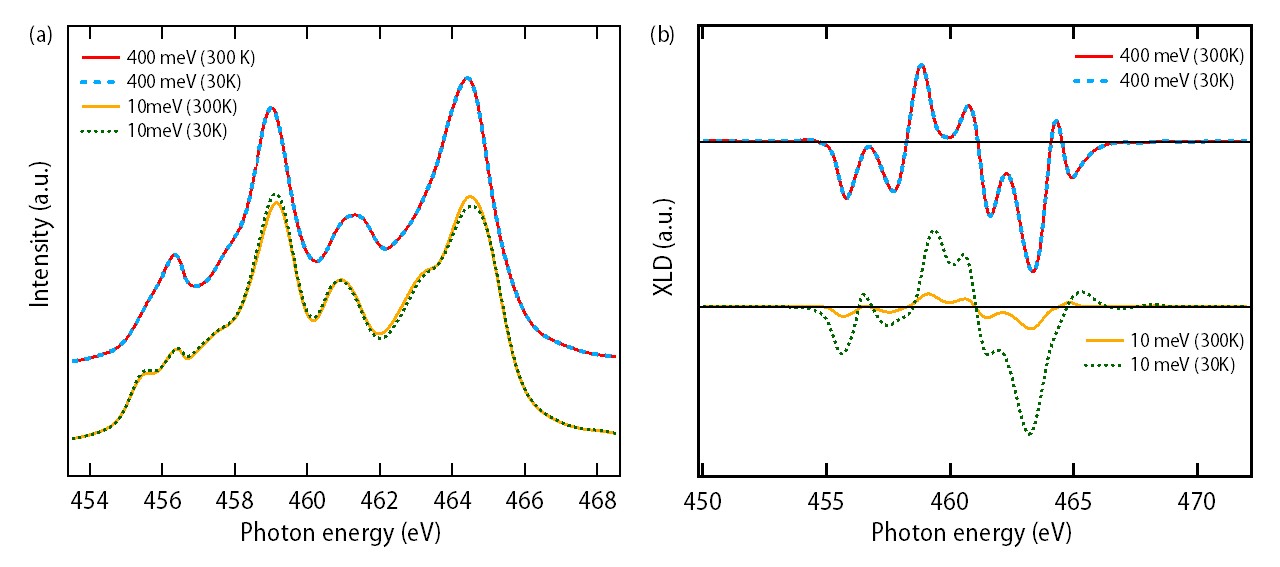}
	\caption{ \textbf{Temperature dependent XAS and XLD simulation for different strength of $D_{\mathrm{2h}}$ distortion.} \label{xld_temp_dep} { (a) and (b), XAS and XLD spectra simulated at 30 K ($\approx$ 2.6 meV) and 300 K (($\approx$ 26 meV)) for $\Delta_1$ = 10 meV and 400 meV.  }}
\end{figure}

To justify the temperature dependence of the XAS and XLD spectra in presence of large orthorhombic distortion, we have performed cluster calculations at 30 K and 300 K for small (10 meV) and large (400 meV) value of  $\Delta_1$. Fig.~\ref{xld_temp_dep} (a) and (b) clearly show that the spectral shape is temperature dependent only when the gap between the lowest occupied state and the next unoccupied state ($\Delta_1$) is small. It is due to the fact that at low temperature  the thermal energy (30 K $\approx$ 2.6 meV) is lower than the gap, however, at high temperature, the thermal energy (300 K $\approx$ 26 meV) is sufficient to excite the carriers from the lowest occupied to the next unoccupied state. Thus, in case of small gap, the thermal occupation changes as a function of temperature, which should not occur when the gap is large. Indeed we observe no temperature dependence of the XAS and the XLD spectra simulated for $\Delta_1$ = 400 meV. We further note, since XLD is an orbital sensitive technique, the change in thermal occupation is reflected in the XLD spectra more prominently than the XAS spectra. Therefore, our observation of temperature insensitivity in the experimental XAS and XLD spectra is consistent with a large orthorhombic distortion.

\clearpage

\begin{flushleft}
	\textbf{S11. Ferro orbital order in strained PTO thin film }
\end{flushleft}

\begin{figure}[h]
	\vspace{-0pt}
	\includegraphics[width=0.7\textwidth] {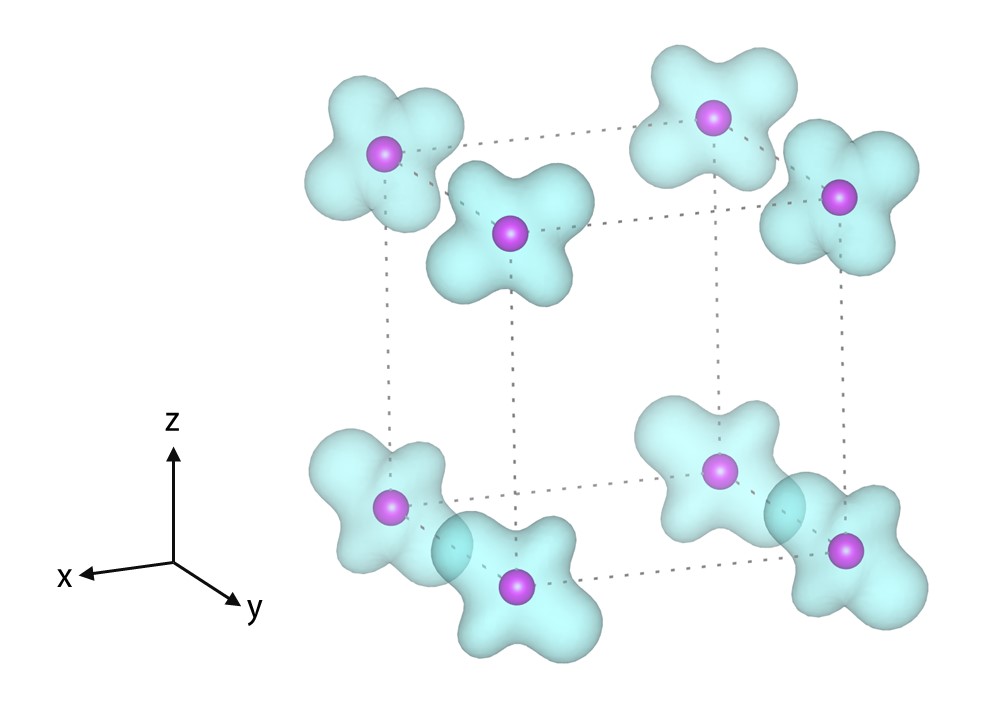}
	\caption{ \textbf{ Ordered pattern of the occupied charge density obtained from DFT.} \label{PTO_FOO} { Isosurfaces of the occupied charge density shown at the Ti (purple) sites. }}
\end{figure}

Our XLD results clearly demonstrates that large orthorhombic distortion ($\approx$ 400 meV) gives rise to a nondegenerate lowest occupied orbital that favors FOO and such ordering should be robust against orbital fluctuations. To find the orbital ordered pattern, we have performed DFT calculation on a strained PTO structure. The constant energy isosurfaces of occupied states and their ordered pattern is shown in Fig.~\ref{PTO_FOO}. The shape of the orbitals is very similar to the previous reports on LaTiO$_3$~\cite{Pavarini:2004p176403, Varignon:2017p235106} and their orientation with respect to the lattice demonstrates the ferro orbital ordering.

\clearpage

\begin{flushleft}
	\textbf{S12. Effect of strain on the orbital ordering }
\end{flushleft}

The orbital physics of antiferromagnetic PrTiO$_3$ thin film, driven by orthorhombic distortion, is entangled with the effect of epitaxial strain (1.6$\%$ compressive). To identify the dominant contribution, we have performed DFT calculations on bulk and strained PTO structures. We find that in both the cases, most occupied orbital is d$_{xz}$ and the occupation of the d$_{yz}$ and d$_{xy}$ changes by small amount suggesting that the effect of orthorhombic distortion is dominant. Thus the orbital symmetry in the FOO state remains unchanged even in the presence of epitaxial strain. In the case where the effect of epitaxial strain dominates the orthorhombic distortion, the TiO$_6$ octahedra should elongate significantly in the out of plane direction (tetragonal distortion) that gives rise to a $D_{4h}$ symmetry around Ti. In case of $D_{4h}$ symmetry, our multiplet calculation shows that the XLD spectra does not agree with the experimental data (see main text Fig. 3), further supporting that the effect of epitaxial strain is minimal in our case.

\begin{figure}[h]
	\vspace{-0pt}
	\includegraphics[width=0.65\textwidth] {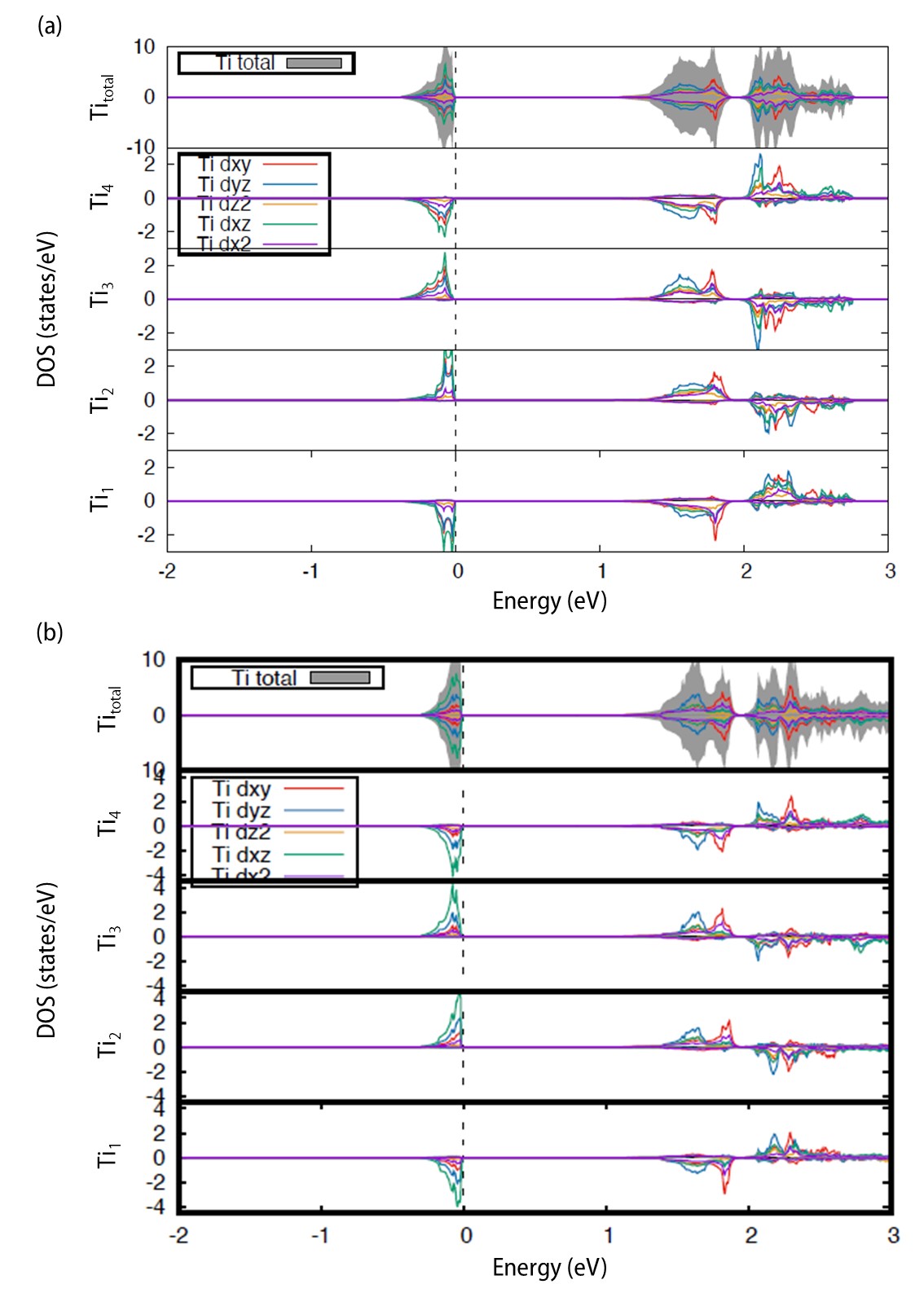}
	\caption{ \textbf{ DOS plots for bulk and strained PTO.} \label{DFT_strain} { (a) and (b) shows the total DOS and its projection on each of the inequivalent Ti sites (See Fig.~\ref{pseudocubic} of SM) in the local TiO$_6$ basis for the bulk and strained PTO structures respectively.}}
\end{figure}

\clearpage

\begin{flushleft}
	\textbf{S13. Orbital occupation in different magnetic configurations }
\end{flushleft}

\begin{figure}[h]
	\vspace{-0pt}
	\includegraphics[width=1.0\textwidth] {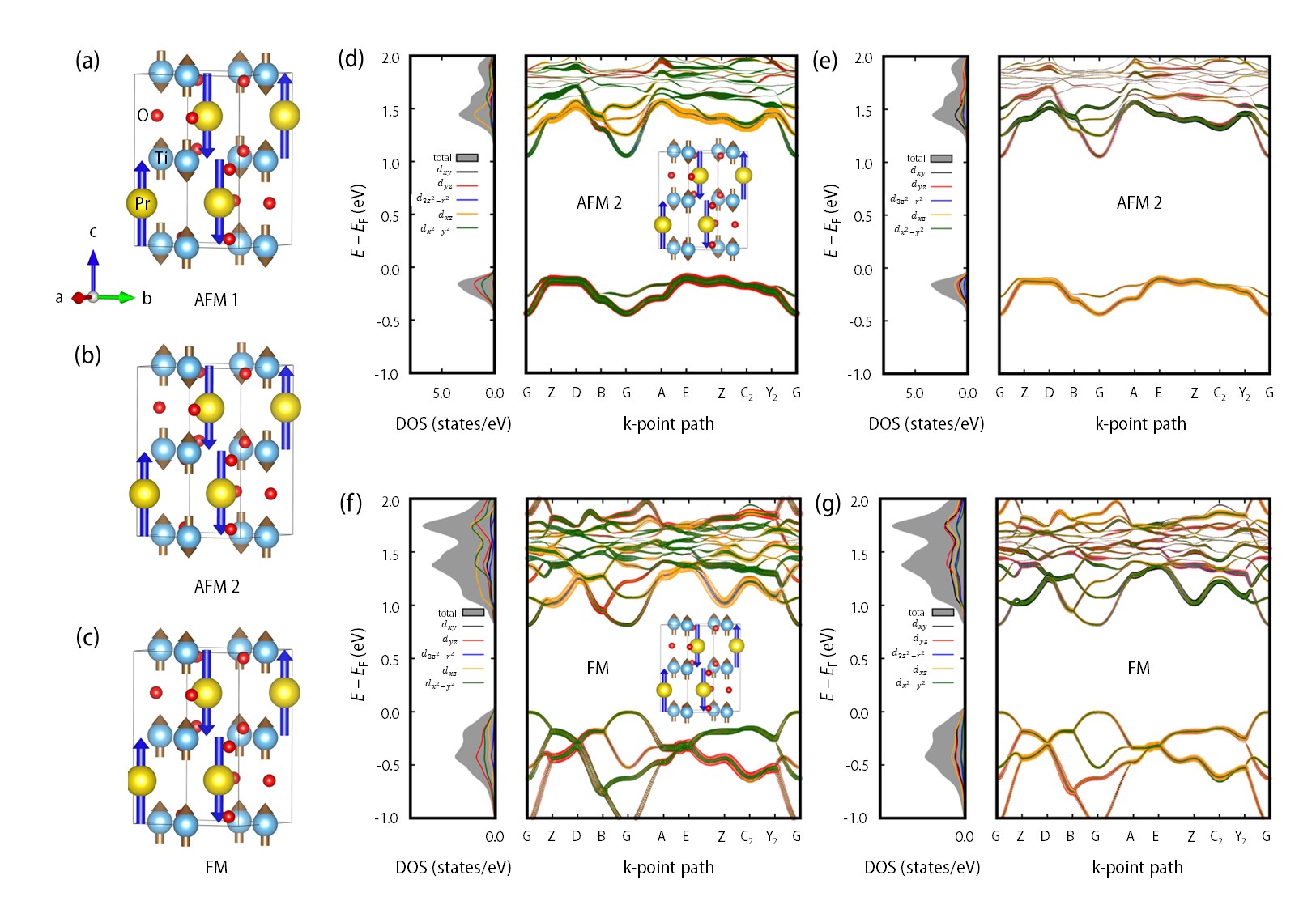}
	\caption{ \textbf{ Electronic structure calculation for different magnetic state} \label{mag_conf} { (a)-(c) The two G-type antiferromagnetic (AFM 1 and AFM 2) and a ferromagnetic (FM) configurations used for the band structure calculations, respectively. The arrows indicate the directions of the magnetic moments. The magnetic moments on Pr atoms are always G-type antiferromagnetic, whereas the moments on Ti atoms are either G-type antiferromagnetic (AFM 1 and AFM 2) or ferromagnetic (FM). (d), (e) and (f), (g) Calculated electronic band structure and density of states for the AFM 2 and FM configurations, respectively. The data is plotted in the global basis ((d) and (f)) and TiO$_{6}$ local basis ((e) and (g)). Black, red, blue, yellow, and green curves (dots) represent $d_{xy}$, $d_{yz}$, $d_{3z^2-r^2}$, $d_{xz}$, $d_{x^2-y^2}$ orbitals, respectively. The Fermi level is set to zero.}}
\end{figure}

In order to get a complete picture of the orbital ordering for the structures with different magnetic configurations (AFM 1 in Fig.~\ref{mag_conf} (a), AFM 2 in Fig.~\ref{mag_conf} (b) and FM in Fig.\ref{mag_conf} (c)), we performed additional first-principles DFT simulations for the other two magnetic states (AFM 2 and FM), in addition to the AFM 1 state which is already discussed in the manuscript. The electronic structure of these two magnetic configurations, are plotted in both the global Cartesian basis (Fig 13 (d) and (f)) and the local TiO$_6$ basis (Fig 13 (e) and (g)). We find that orbital occupation does not change with magnetic configuration, i.e., the dominant $d_{xz}$ occupation just below the Fermi level is robust regardless of the magnetic state. Different magnetic states induce slight modifications in the hybridization between Ti-$d$ electrons and O-$p$ electrons, leading to an increase in the occupation of $d_{yz}$ and $d_{xy}$. However, $d_{3z^2-r^2}$ and $d_{x^2-y^2}$ remain degenerate, with almost zero occupation. Furthermore, the dispersive bands observed in the FM state (Fig.\ref{mag_conf} (c)) can be explained by the fact that the FM state is known to give $d$ electrons a more itinerant character. Thus we find that while varying magnetic states can have a minor impact on the electronic structure, the $d_{xz}$ orbital occupation remains unaffected by such changes signifying the dominant effect orthorhombic distortion on the orbital ordering.

\end{document}